\documentclass[twocolumn]{aastex61}
\usepackage{tabularx}
\received{}
\revised{}
\accepted{ApJS \today }

\shorttitle{SPS of star forming clumps in galaxy pairs}
\shortauthors{Zaragoza-Cardiel et al.}

\begin{document}

\title{Stellar Population Synthesis of star forming clumps in galaxy pairs and non-interacting spiral galaxies}

\correspondingauthor{Javier Zaragoza-Cardiel}
\email{javier.zaragoza@astro.unam.mx, javier.zaragoza@um.es}

\author[0000-0001-8216-9800]{Javier Zaragoza-Cardiel}
\affiliation{Instituto de Astronom\'ia, \\ 
Universidad Nacional Aut\'onoma de M\'exico, \\ 
04510, D. F., M\'exico \\
}

\author{Beverly J. Smith}
\affiliation{Department of Physics \& Astronomy \\
East Tennessee State University, \\
Johnson City, TN, USA}

\author{Margarita Rosado}
\affiliation{Instituto de Astronom\'ia, \\ 
Universidad Nacional Aut\'onoma de M\'exico, \\ 
04510, D. F., M\'exico \\
}

\author{John E. Beckman}
\affiliation{Instituto de Astrof\'isica de Canarias, \\
C/ V\'ia L\'actea s/n, \\
38200 La Laguna, Tenerife, Spain\\}
\affiliation{Departamento de Astrof\'isica, \\
Universidad de La Laguna, \\
Tenerife, Spain\\}
\affiliation{Consejo Superior de Investigaciones Cient\'ificas, \\
Spain\\}

\author{Theodoros Bitsakis}
\affiliation{CONACYT Research Fellow - Instituto de Radioastronom\'ia y Astrof\'isica, \\
Universidad Nacional Aut\'onoma de M\'exico, \\
58190 Morelia, M\'exico\\}

\author{Artemi Camps-Fari\~{n}a}
\affiliation{Instituto de Astrof\'isica de Canarias, \\
C/ V\'ia L\'actea s/n, \\
38200 La Laguna, Tenerife, Spain\\}
\affiliation{Departamento de Astrof\'isica, \\
Universidad de La Laguna, \\
Tenerife, Spain\\}
\affiliation{Instituto de Ciencias Nucleares, \\
Universidad Nacional Aut\'onoma de M\'exico, \\ 
04510, D. F., M\'exico \\}

\author{Joan Font}
\affiliation{Instituto de Astrof\'isica de Canarias, \\
C/ V\'ia L\'actea s/n, \\
38200 La Laguna, Tenerife, Spain\\}
\affiliation{Departamento de Astrof\'isica, \\
Universidad de La Laguna, \\
Tenerife, Spain\\}

\author{Isaiah S. Cox}
\affiliation{Department of Physics \& Astronomy \\
East Tennessee State University, \\
Johnson City, TN, USA}



\begin{abstract}
We have identified $1027$ star forming complexes in a sample of 46 
galaxies from the Spirals, Bridges, and Tails (SB\&T) sample of interacting galaxies, and 693 star forming complexes in a sample of 38 
non-interacting spiral (NIS) galaxies in $8\rm{\mu m}$ observations from 
the {\it Spitzer} Infrared Array Camera. We have used archival 
multi-wavelength UV-to IR observations to fit the observed spectral energy distribution of our clumps with 
the Code Investigating GALaxy Emission using a 
 double exponentially declined star formation history. We derive the star formation rates 
(SFRs), stellar masses, ages and fractions of the most recent burst, dust attenuation, and fractional emission due to an active galactic nucleus 
for these clumps. The resolved star formation main sequence holds on 2.5kpc scales, although it does 
not hold on 1kpc scales.
 We analyzed the relation between SFR, stellar mass, and age of the recent burst in the SB\&T and NIS samples, and  
we found that the SFR per stellar mass is higher in the SB\&T galaxies, and the clumps are younger
 in the galaxy pairs.  We analyzed the 
 SFR radial profile and found that SFR is enhanced through the disk and in the tidal features relative
to normal spirals.

\end{abstract}

\keywords{}



\section{Introduction} \label{sec:intro}

Galaxy mergers are key ingredients of galaxy mass growth and morphological transformation in the hierarchical scenario of galaxy formation 
within the  standard cosmological model \citep{2005Natur.435..629S,2006ApJ...645..986R,2011EAS....51..107B}. Moreover, they were more common at 
higher redshifts, therefore local galaxy mergers are often used as nearby analogs to improve our understanding of the phenomena involved in high redshift galaxy evolution. 

Since \citet{1978ApJ...219...46L} showed evidence of a burst mode of star formation in peculiar galaxies, several studies have found that galaxy 
interactions can enhance star formation rates by a factor of 2-3 on average relative to their stellar mass 
\citep{1987ApJ...320...49B,1987AJ.....93.1011K,2007AJ....133..791S,2007ApJ...660L..51L,2008MNRAS.385.1903L,2015MNRAS.454.1742K}.
In fact, the most intense star forming 
galaxies in the nearby Universe, the Ultra Luminous Infrared Galaxies, are mostly driven by mergers \citep{1998ApJS..119...41K}. 
One might expect that the most intense star forming galaxies at the peak of the cosmic star formation rate, $z\sim2$ 
\citep{2014ARA&A..52..415M} would be driven by mergers. However, even using the same data different 
authors reach different conclusions \citep{2015ApJ...799..209W,2017MNRAS.465.1157R} due to differences in the merger classification criteria. 

Resolved star formation studies of nearby interacting galaxies are crucial to identify which processes are enhancing the star formation. 
 Simulations show that galaxy mergers can produce
a loss of axisymmetry, producing gas flows toward the central parts
of the galaxies \citep{1996ApJ...464..641M}, and therefore central starbursts \citep{2007A&A...468...61D}.
 However, more recent high resolution simulations also produced 
extended star formation due to shock-induced
star formation \citep{2004MNRAS.350..798B,2010MNRAS.407...43C}, or enhanced compressed modes of turbulence \citep{2011EAS....51..107B,2013MNRAS.434.1028P,2014MNRAS.442L..33R}.
Analytical models show that tidal disturbances between galaxies perturb the orbits of interstellar clouds, 
producing high density orbiting crossing zone zones
in the outer disks and tidal tails, presumably triggering star formation \citep{2012MNRAS.422.2444S}.  
However, smoothed particle hydrodynamical (SPH) simulations of pre-merger interacting pairs run by \citet{2015MNRAS.448.1107M}
find suppressed star formation at radii greater than 1 kpc, compared to isolated galaxies.

Observationally, off-nuclear enhanced star formation is seen in individual cases 
\citep{1978IAUS...77..279S,2004MNRAS.350..798B,2004ApJS..154..193W,2007AJ....133..791S,2010MNRAS.407...43C,2010AJ....139.1212S}.  
Larger samples of interacting galaxies are needed to obtain better statistical 
information on how mergers affect star formation and therefore galaxy evolution. 
\citet{2016AJ....151...63S} presented the analysis of $\sim700$ star forming regions in a sample of 46 galaxy pairs and compared them with 
those of regions in a sample of 39 normal spiral galaxies, showing that 
the SFR is proportionally higher for the star forming regions in galaxy pairs.
\citet{2015MNRAS.451.1307Z} found an  
enhancement in electron density, SFR, and velocity dispersion of $\sim1000$ H{\sc ii} regions in galaxy pairs compared to $\sim1000$ H{\sc ii} regions in non-interacting spirals,  
analyzing H$\alpha$ emission, 
consistent with the picture of higher gas turbulence, and higher massive star formation induced by mergers \citep{2011EAS....51..107B}. Nevertheless,   
neither dust attenuation nor stellar population were analyzed in \citet{2015MNRAS.451.1307Z}, since their main purpose was 
the study of the internal kinematics of H{\sc ii} regions with very high spectral resolution.

Stellar population synthesis can be used to obtain the contribution of the interaction to the star formation in terms of the age of 
the stellar population, and the star formation rate compared to the stellar mass.
A well-defined relationship between the global SFR of star-forming galaxies
and their stellar mass, $\rm{M_{*}}$, has been discovered \citep{2004MNRAS.351.1151B,2007ApJS..173..267S}; 
this is known as
the star formation main sequence of galaxies.
This main sequence evolves with redshift out to $z \sim 6$ \citep{2007ApJ...670..156D,2009MNRAS.393..406C}, 
but at a given redshift, the scatter in the SFR for a given stellar
mass is consistent at $\sim 0.2 \thinspace\rm{dex}$ \citep{2014ApJS..214...15S}.
Recently, \citet{2016ApJ...821L..26C} found that the star formation main sequence 
still holds on $\rm{kpc}$ scales in a sample of 306 galaxies from The Calar Alto Legacy Integral Field Area survey (CALIFA; \citet{2012A&A...538A...8S}), claiming 
that the star formation process is mainly a local process rather than a global one.  Similar recent studies concluded that the resolved star formation main sequence 
holds on kiloparsec scales in nearby galaxies \citep{2017MNRAS.466.1192M,2017MNRAS.469.2806A} and at redshift $z\sim1$ \citep{2013ApJ...779..135W,2016MNRAS.456.4533M}.
performed spatially resolved population synthesis for nine galaxy pairs, and found younger stellar 
populations than those seen in isolated galaxies.  They concluded that this was due to gas flows caused by the interaction.

In the current study we present a stellar population synthesis analysis of the \citet{2016AJ....151...63S} regions  
using UV, optical, and IR observations. We then construct the resolved main sequence for the two samples of galaxies and investigate 
the SFR per stellar mass, the ages of the stellar component,  and the spatial extent of the SFR in galaxy pairs.
In section $\S2$ we briefly present the samples, and the photometry of the star forming complexes that were already presented in \citet{2016AJ....151...63S}. 
In section $\S3$ we describe the method used to fit the spectral energy distributions (SEDs) of the clumps to model SED. In section $\S4$ we present the results of the SED analysis, while 
in section $\S5$ we show the analysis of the SFR radial variation. Finally, in $\S6$ we give a discussion and draw our conclusions.

\section{Data \& clump photometry }

\subsection {Data}
We have previously presented in \citet{2016AJ....151...63S} the identification of $\sim 700$ star forming complexes in galaxies from the Spirals, Bridges, and Tails (SB\&T) sample \citep{2007AJ....133..791S,2010AJ....139.1212S}, 
and star forming complexes in a control sample of non-interacting spiral (NIS) galaxies obtained from \citet{2003PASP..115..928K,2007ApJS..173..185G}. 
 We present both samples in Tabs. \ref{tab_sbtsample} and \ref{tab_nissample}.
 The SB\&T sample is composed of pre-merger galaxies pairs chosen from the Arp Atlas \citep{1966apg..book.....A}, 
with velocities $<10,350 \rm{km/s}$ and angular sizes $\gtrsim3\arcmin$, plus NGC 4567/8 and NGC 2207/IC 2163 that are not in 
the Arp Atlas. The total S\&BT sample has 46 pairs, while there are 38 NIS.


The data we used for this study include the GALEX NUV and FUV, {\it Spitzer} IRAC
3.6$\mu\rm{m}$, 4.5$\mu\rm{m}$, 5.8$\mu\rm{m}$, 8.0$\mu\rm{m}$, and {\it Spitzer} MIPS 24$\mu\rm{m}$
data used in \citet{2016AJ....151...63S}.
For the current study, for the 37 out of 46 galaxy pairs, and the 31 out of 38 spirals with optical Sloan Digitized Sky Survey (SDSS) images we used those data as well.  
The SDSS $\textit{ugriz}$ filters have effective wavelengths 
of 3560 $\rm{\AA}$, 4680 $\rm{\AA}$, 6180 $\rm{\AA}$, 7500 $\rm{\AA}$, and 8870 $\rm{\AA}$ respectively. 
The SDSS FWHM spatial resolution is typically about $1.3\arcsec$.   
For all of the galaxies in the sample, we also carried out clump photometry using the J, H, and K$_{\rm S}$ maps from the 2MASS survey.  
These bands have effective wavelengths of 1.25 $\mu$m, 1.65 $\mu$m, and 2.17 $\mu$m respectively. These images have a 
spatial resolution of $\sim4\arcsec$ \citep{2006AJ....131.1163S}.

To determine total fluxes for the sample galaxies in these filters,
we used a set of rectangular boxes that covered the observed extent
of the galaxy in the images, but avoided very bright stars. 
These regions included all of the clumps identified in the tidal
features (see below for the identification and classification of the clumps).
For each image, the sky was determined using rectangular sky regions
off of the galaxies without bright stars or other sources.
Total fluxes for the individual galaxies in a pair were determined 
separately and treated separately in the analysis.

\subsection{Identification of clumps}

We have identified the clumps in smoothed 8 $\mu\rm{m}$ observations from the {\it Spitzer} Infrared Array Camera \citep{2004ApJS..154...10F}.  
Although the 24 $\mu\rm{m}$ filter is considered a better tracer of star formation than 8 $\mu\rm{m}$ 
(e.g., \citet{2005ApJ...633..871C,2007ApJ...666..870C}), 
{\it Spitzer} 24 $\mu\rm{m}$ images suffer from more artifacts, and have lower native spatial resolution than the 8 $\mu\rm{m}$ band.  
The 8 $\mu\rm{m}$ band is also a better choice than H$\alpha$ to identify star forming regions in our sample, since our H$\alpha$ dataset 
is incomplete and inhomogeneous, and the H$\alpha$ is strongly affected by dust absorption.  
The UV bands also suffer from extinction, thus a clump search on UV maps may miss the most obscured regions in interacting 
galaxies and may produce positions that are offset from the peak of the star formation \citep{2014AJ....147...60S}.

\begin{figure*}

\gridline{\fig{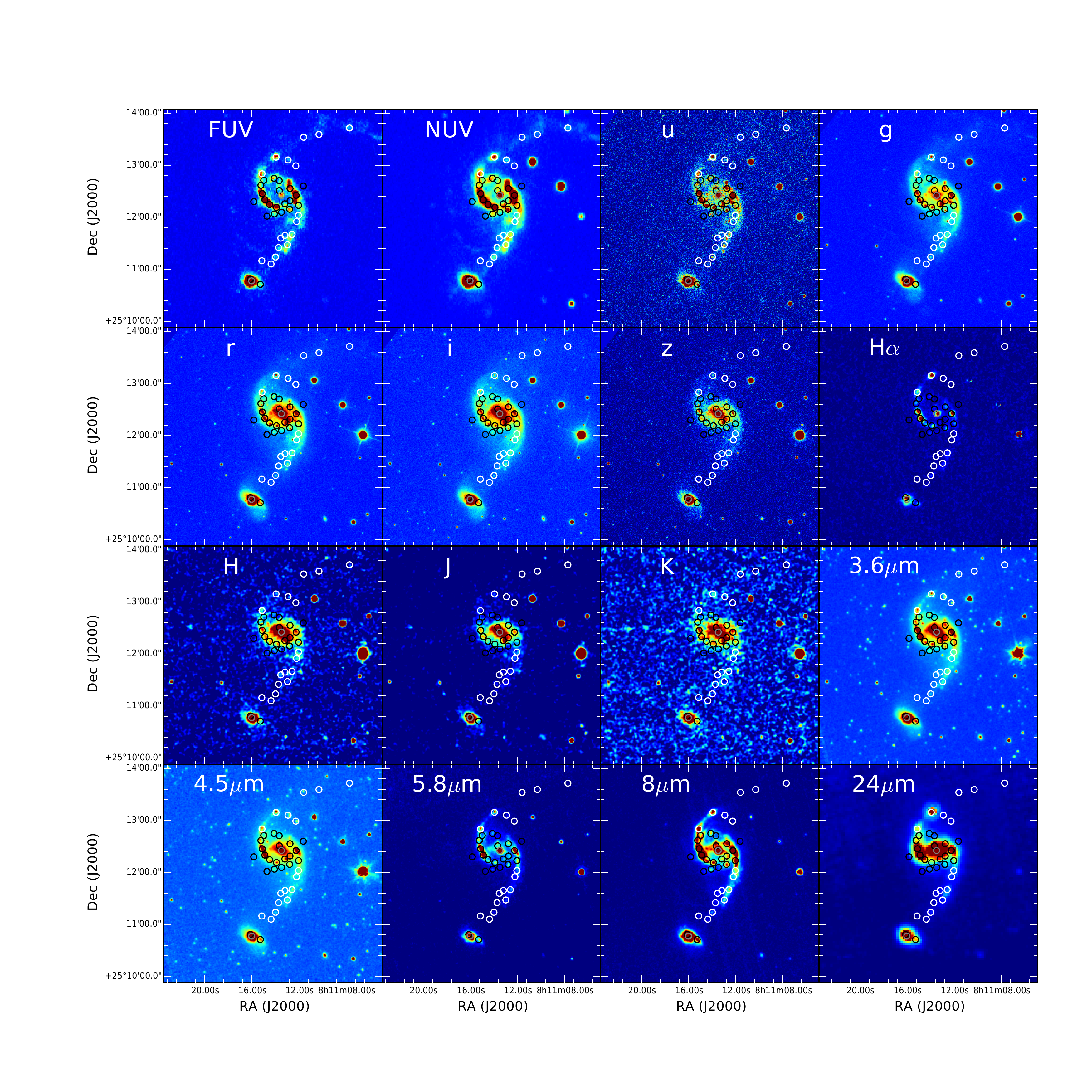}{0.6\textwidth}{(a)}}
\gridline{\fig{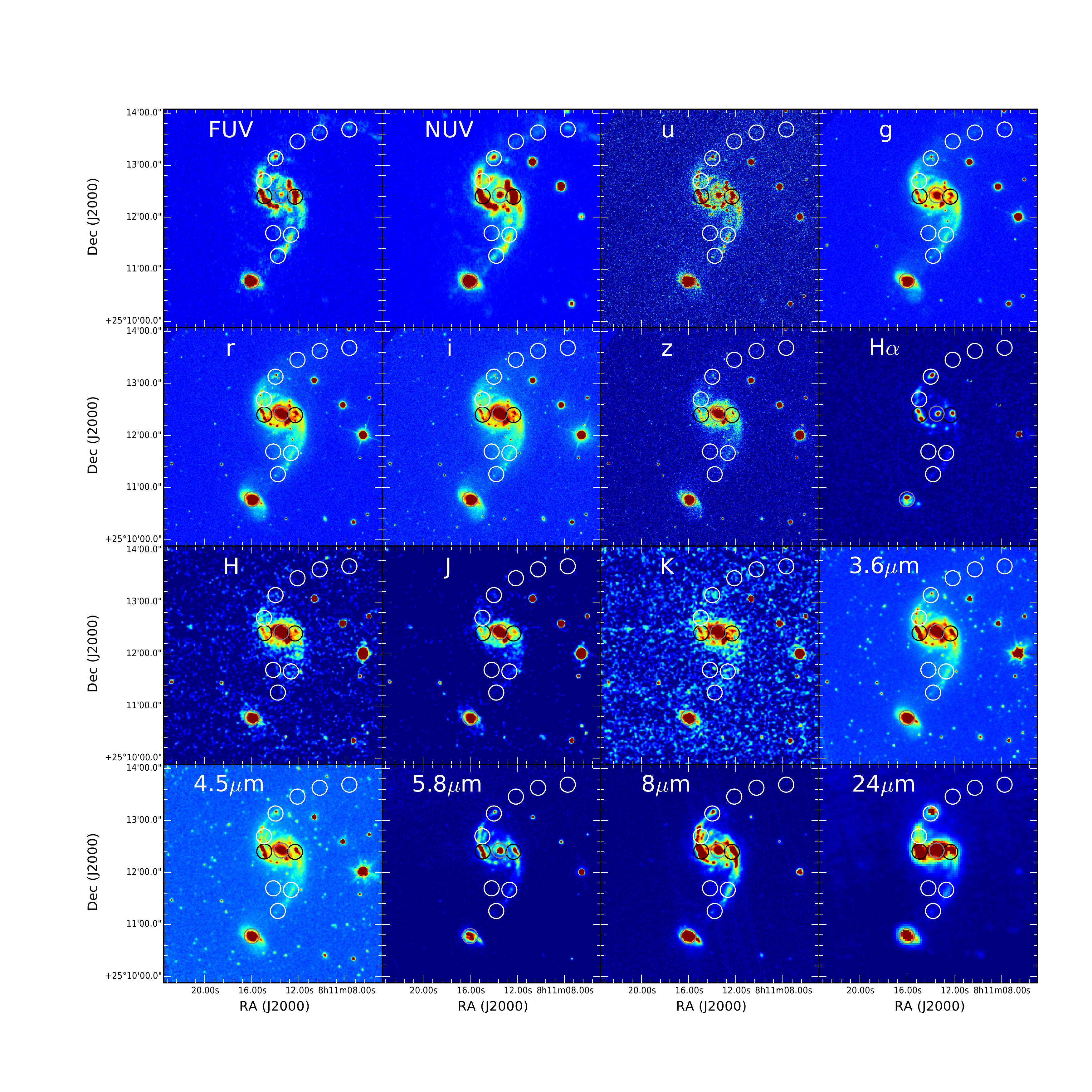}{0.6\textwidth}{(b)}}


\caption{GALEX, SDSS, H$\alpha$, 2MASS, and {\it Spitzer} images of Arp 82. (a) Identified clumps at 1kpc scales. (b) Identified clumps at 2.5 scales. 
Black circles are clumps in the disk, white are those in the tails, and gray are those in the nucleus. 
\label{fig_clumps_arp82}
}
\end{figure*}

For the identification of clumps, two different Gaussian smoothings were used, one that produces a FWHM resolution of 1 kpc, 
and the other of 2.5 kpc.   As described in detail in \citet{2016AJ....151...63S}, clumps were selected automatically from the smoothed 
images using the Image Reduction and Analysis Facility (IRAF) \footnote{\href{http://iraf.noao.edu}{http://iraf.noao.edu}} daofind routine \citep{1987PASP...99..191S} 
using a detection threshold of 10 sigma above the noise level.  The daofind parameters sharplo, sharphi, roundlo, and roundhi 
were set to 0.1, 1.2, -2.0, and 2.0, respectively, to allow slightly extended and/or elongated clumps.  The images were then 
inspected visually, to eliminate spurious detections due to artifacts in the images. We show in Fig. \ref{fig_clumps_arp82} the identified 
clumps: (a) 1kpc, (b) 2.5kpc; for Arp 82 in all the observed bands. 

\subsection{Photometry of the clumps}

The photometry of the clumps was then performed on the unsmoothed images using the IRAF daophot routine with aperture radii of 1.0 and 2.5 kpc, 
respectively.  The local galaxian background was calculated using a sky annulus with an inner radius equal to the aperture radius, and an 
annulus width equal to 1.2 $\times$ the aperture radius.  The mode sky fitting algorithm was used to calculate the background level, 
as the mode is considered most reliable in crowded fields \citep{1987PASP...99..191S}.  The poorer spatial resolution in the GALEX bands and at 24 
$\mu\rm{m}$ may lead to greater clump contribution to the sky background, and so slightly lower fluxes.

The fluxes were then aperture-corrected to account for spillage outside of the aperture due to the image resolution.   
For the GALEX, 2MASS, and SDSS images, the aperture corrections were calculated for each image individually.  
For each image, aperture photometry for three to ten moderately bright isolated point sources was done using our target aperture radius, 
and then comparing with photometry done within a $17\arcsec$ radius.
More details on this process are provided in Smith et al. (2016).
For the {\it Spitzer} data, rather than calculating aperture corrections ourselves we interpolated between the tabulated values of 
aperture corrections provided in the IRAC and MIPS Instrument Handbooks 
\footnote{ http://irsa.ipac.caltech.edu/data/SPITZER/docs/ }.
We were not able to calculate aperture corrections for the H$\alpha$ fluxes 
because of the lack of isolated off-galaxy 
point sources on the H$\alpha$ maps. The aperture corrections for H$\alpha$ are
expected to be small because of the relatively high spatial resolution 
($0.7\arcsec$ to $1.5\arcsec$).
If the intrinsic size of a clump is
large compared to our aperture radii,
our aperture corrections (which assume point sources)
may underestimate the true fluxes, particularly for bands with poor
intrinsic resolution.
For example, some of the clumps may
be blends of multiple smaller clumps, with one of our clumps
consisting of several smaller components.  Alternatively,
a clump may be a single physically-large object.   In these cases, 
our final fluxes in the filters with lowest resolution 
(GALEX and {\it Spitzer} 24 $\mu\rm{m}$), 
may be somewhat under-estimated compared to filters with
better spatial resolution.

We used the $1\rm{kpc}$ radii clumps to study star formation 
on smaller scales for the 30 galaxy pairs and 36 NIS galaxies closer than $67\thinspace \rm{Mpc}$, and used the $2.5\rm{kpc}$ radii clumps to study  star formation 
on a larger scale in the whole sample; 
$2.5\rm{kpc}$ is the limiting resolution ($6\arcsec$ FWHM in GALEX and {\it Spitzer} $24\rm{\mu m}$).
This choice of parameters allowed us to obtain accurate photometry even in the furthest galaxy, Arp107 at $142\rm{Mpc}$. 

 


In Table \ref{tableclumps_photometry} we present the photometry for GALEX: NUV and FUV; IRAC: $3.6\rm{\mu m}$, $4.5\rm{\mu m}$, $5.8\rm{\mu m}$, $8.0\rm{\mu m}$; MIPS $24\rm{\mu m}$; SDSS: 
u, g, r, i, z; H$\alpha$+cont, continuum subtracted H$\alpha$, and 2MASS: J, H, K. Three different classifications for the clumps in the SB\&T sample were used  
as explained in \citet{2016AJ....151...63S}: clumps in the disk, 
in tails, and in the nuclear region; for the clumps in the NIS sample we classified the clumps in the disk, and those in nuclear regions.  
Thus, the column containing the name of the clumps consists of the name of the system (galaxy in the case of NIS galaxies), consecutive identification number, 
the sample to which it belongs, location, and radius of the aperture 
in kiloparsec. In the fourth column of Table \ref{tableclumps_photometry}, we provide the galaxy name; for the SB\&T
galaxies, this is the name of the individual galaxy in the pair the clump is associated with.

\section{SED modeling}

\begin{table*}
{\scriptsize
\caption{CIGALE parameters\label{tab_cig}}
 \centering
 \begin{tabularx}{\textwidth}{lX}
 \hline
  Free parameters&\\\hline\hline
  $e$-folding time of the old population& 2, 4, 6 Gyr\\
  $e$-folding time of the late starburst population & 5, 10, 25, 50, 100 Myr\\
  Mass fraction of the late burst population& 0, 0.2, 0.4, 0.6, 0.8, 0.99\\
  Age of the late burst&1, 5, 10, 15, 20, 25, 30, 40, 50, 60, 70, 80, \\&   90, 100, 150, 200, 250, 300, 400, 500 Myr\\
  Metallicity& 0.008, 0.02, 0.05\\
  E(B$-$V) of the stellar continuum light for the young population.&0.01,  0.2,   0.4, 0.6, 0.7 mag\\
  Amplitude of the UV bump&0, 1, 2, 3 \\  
  Slope $\delta$ of the power law modifying the attenuation curve&-0.5, -0.3, -0.1, 0.0\\
  AGN fraction (just for nuclear regions) & 0.0,0.2,0.4,0.6,0.8,0.99 \\
  $\alpha$ slope&1.0, 1.5, 2., 2.5, 3., 3.5, 4.\\\\\hline

  Fixed parameters&\\\hline\hline
  Age of the oldest stars&13 Gyr\\
  Reduction factor for the E(B-V) of the old population compared to the young one &0.44\\
    IMF & \citet{2003PASP..115..763C}\\
  Ionization parameter&$10^{-2}$\\
  Fraction of Lyman continuum photons absorbed by dust&10\%\\
  Fraction of Lyman continuum photons escaping the galaxy&0\%\\\hline
 \end{tabularx}
 }
\end{table*}

We use the Code for Investigating GALaxy Emission\footnote{\href{http://cigale.lam.fr}{http://cigale.lam.fr}} (CIGALE, \citet{2009A&A...507.1793N}), python version 0.9,  to model 
and fit the SEDs for each individual clump. 

CIGALE is based on the assumption of an energy balance between the energy 
absorbed in the UV, optical, and NIR, and re-emitted by the dust in the MIR and FIR. CIGALE uses the dust emission model of \citet{2014ApJ...784...83D} which is 
dependent on the relative contribution of different heating intensities, $U$, modeled by the exponent $\alpha$ in the spatially integrated 
dust emission $d M_{d}\propto U^{\alpha}dU $, where $M_d$ is the dust mass heated by a radiation intensity $U$ \citep{2002ApJ...576..159D}. 
We leave $\alpha$ as a free parameter, and for the nuclear regions we also leave the AGN fraction contribution as a variable  (see Table \ref{tab_cig}) while 
for the rest of the clumps we set the AGN fraction contribution to zero. To model dust attenuation, CIGALE assumes a 
combination of dust attenuation curves from \citet{2000ApJ...533..682C} and \citet{2002ApJS..140..303L} and modifies them by a power law centered at $550\rm{nm}$, with 
exponent $\delta$ (free parameter), and adds a UV bump with a specific amplitude (free parameter). We fix the differential reddening, and leave the 
color excess E(B-V) as a free parameter.

 In order to model the plausible recent star formation enhancement in galaxy pairs, we model the star formation history with two decaying 
exponentials:

\begin{equation}
 {\rm{SFR}}(t)=(1-f_{\rm{y}})\thinspace{\rm{SFR_{0\thinspace old}}}\thinspace e^{-\frac{t-t_1}{\tau_1}}+f_{\rm{y}}\thinspace{\rm{SFR_{0\thinspace young}}}\thinspace e^{-\frac{t-t_2}{\tau_2}}
 \label{eq_sfh}
\end{equation}

as described in \citet{2011ApJ...740...22S},
where the $e$-folding times ($\tau_i$), the mass fraction of the recent starburst ($f_{\rm{y}}$), and the age ($t_2$) of the recent starburst, are left as free parameters, while the age 
of the oldest stars ($t_1$) is set (see Table \ref{tab_cig} for values).
We use the stellar populations of \citet{2003MNRAS.344.1000B} considering the \citet{2003PASP..115..763C} initial mass function, and three 
possible values of metallicity (around solar). The CIGALE parameters are summarized in Table \ref{tab_cig}.

The aforementioned set of parameters yields $3\cdot 10^{6}$ models for non-nuclear regions and $1.8\cdot 10^{7}$ models for nuclear regions, and then 
CIGALE performs a Bayesian analysis for each output parameter as described in \citet{2009A&A...507.1793N}, 
resulting in the estimated values and uncertainties given in Table \ref{tab_cig_out}. To be sure of the goodness of the fit, 
we include only the clumps for which the fit of the SED yields $\chi^2 _{\rm{red}}<10$.  For those clumps with no SDSS 
observations ($27\%$) the relative uncertainties of the resulting parameters are on average only $2\%$ larger, thus we can include them in the 
analysis directly with the rest of the clumps.
Additionally, since we are interested in the study of 
recent star formation, we do not consider in the following analysis the clumps and galaxies with no present star formation, {\it i. e. }, 
$f_{\rm{y}}=0$.

\subsection{SED vs. photometric star formation rates}

\begin{figure}
\plotone{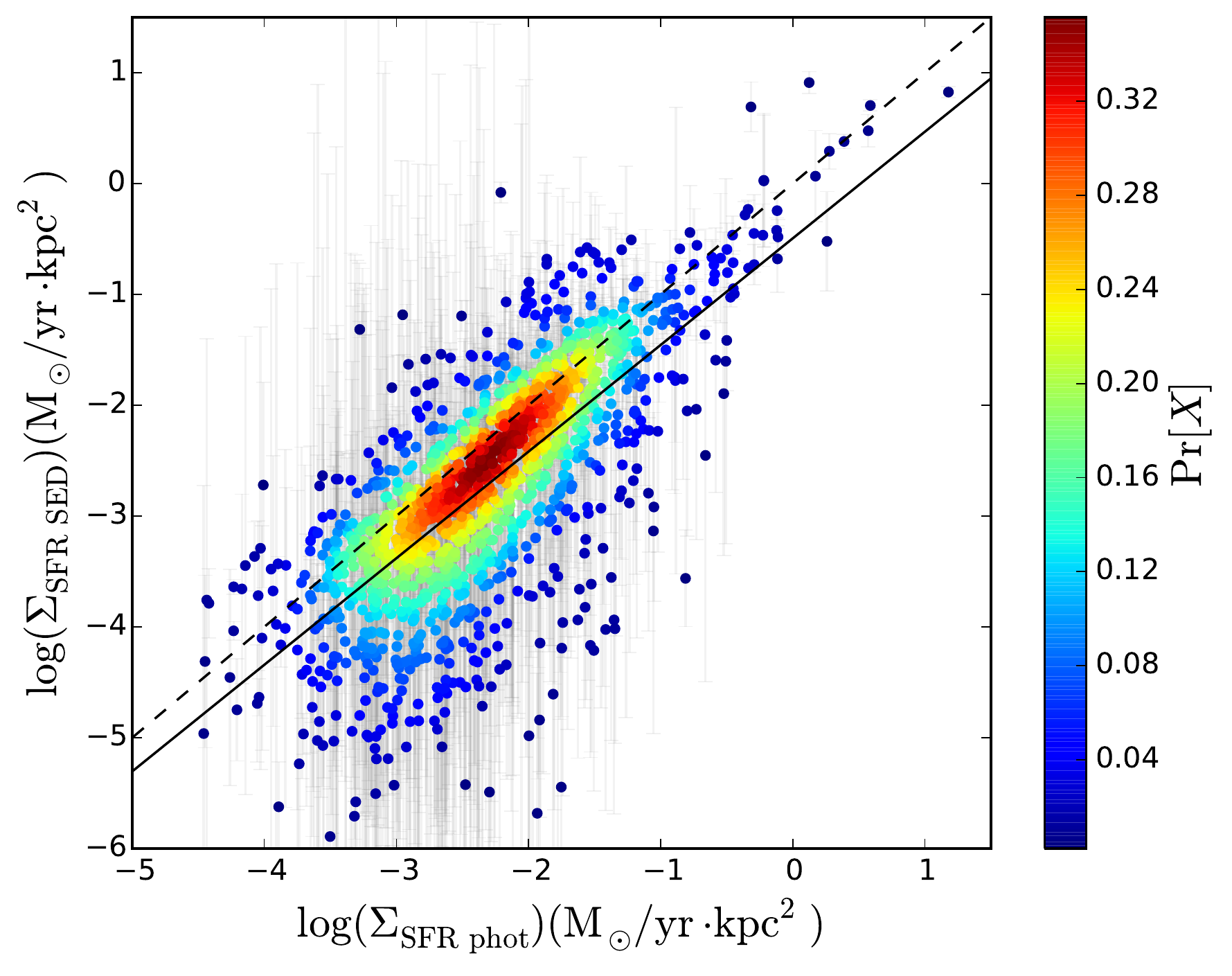}
\caption{
Instantaneous SFR surface density derived from the SED fitting, $\Sigma_{\rm{SFR\, SED}}$, versus the SFR surface density presented in \citet{2016AJ....151...63S}, 
$\Sigma_{\rm{SFR\, phot}}$, 
color coded with the  probability distribution function derived from the data points, ${\rm{Pr}}[X]$.
The solid line represents the variable $x$-bin size linear fit (Eq. \ref{eq_fit_sfr16_cig}), while the dashed line represents the 
one to one relation. 
\label{fig_sfrbest_cig}}
\end{figure}

We plot in Fig. \ref{fig_sfrbest_cig} the instantaneous SFR surface density derived from the SED fitting, $\Sigma_{\rm{SFR\, SED}}$, versus the 
photometric SFR surface density 
derived from UV + IR fluxes, 
$\Sigma_{\rm{SFR\, phot}}$, presented in 
\citet{2016AJ....151...63S}, for all the identified clumps. 
These points are color coded with the  probability distribution function (PDF) derived from the data points, ${\rm{Pr}}[X]$. We use the same 
color code in the later figures of this work where we color coded  with the PDF. 
The variable $x$-bin size linear fit (solid line) yields:

\begin{equation}
 \log(\Sigma_{\rm{SFR\, SED}})= (0.96\pm0.03)\cdot\log(\Sigma_{\rm{SFR\, phot}}) -(0.49\pm0.08).
 \label{eq_fit_sfr16_cig}
\end{equation}

Thus, using the SFR obtained from the SED fitting is equivalent to using the photometric SFR presented in \citet{2016AJ....151...63S}, since they just 
differ in a constant shift compared to the one to one relation (dashed line in Fig. \ref{fig_sfrbest_cig}). We will use in the following analysis the instantaneous 
SFR derived from the SED fitting.

%
%
%

\section{Results}

\subsection{Integrated star formation main sequence}

\begin{figure}
\plotone{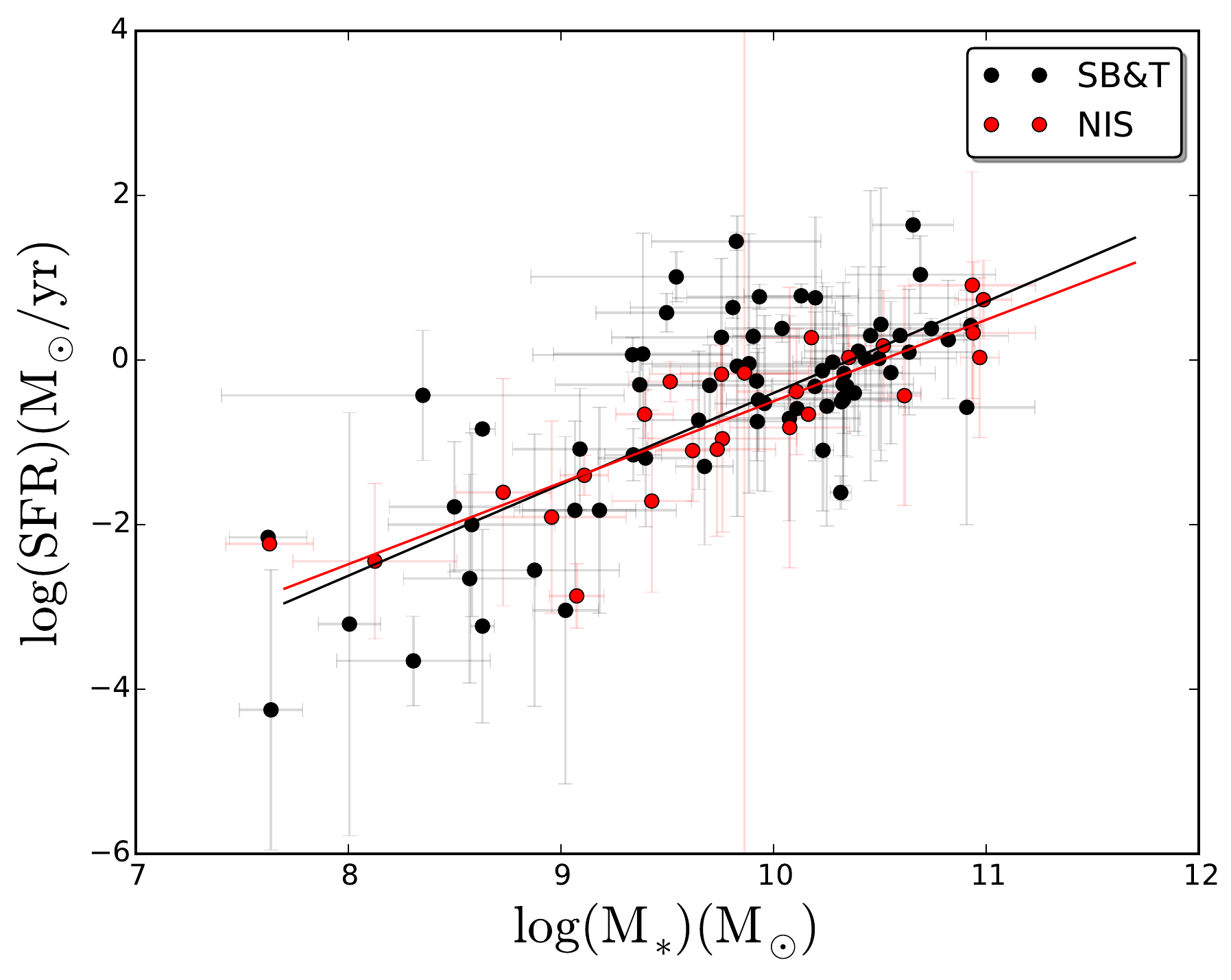}
\caption{Star formation rate, $\rm{SFR}$, versus stellar mass, $\rm{M_{*}}$, for the SB\&T galaxies (black) and NIS galaxies 
(red).
The lines are the linear fits to the data points. 
\label{fig_sfms}}
\end{figure}

We obtained integrated aperture photometry for each galaxy in the SB\&T and NIS sample in the same bands as in the clumps. The integrated photometry is presented 
in Tab. \ref{tableintegrated_photometry}. Then, we used the same 
set of CIGALE parameters (Table \ref{tab_cig}) to derive the integrated SFR and $M_*$ for each galaxy. We plot 
the star formation main sequence, SFR versus $M_*$, in Fig. \ref{fig_sfms} for the SB\&T galaxies in black, and 
the NIS in red. We perform linear fits to both samples separately, and we obtain:

\begin{equation}
 \log(\rm{SFR})\left(\frac{\rm{M_{\odot}}}{\rm{yr}}\right)=(1.11\pm0.13)\log(\rm{M_{*}})\left(\rm{M_{\odot}}\right)-(11.5\pm1.2),
 \label{eq_intsbt}
\end{equation}

for SB\&T galaxies (black line in Fig \ref{fig_sfms}), and

\begin{equation}
 \log(\rm{SFR})\left(\frac{\rm{M_{\odot}}}{\rm{yr}}\right)=(0.99\pm0.12)\log(\rm{M_{*}})\left(\rm{M_{\odot}}\right)-(10.4\pm 1.2),
 \label{eq_intspi}
\end{equation}

for NIS (red line in Fig \ref{fig_sfms}).  The scatter of the integrated star formation main sequence 
after removing the average uncertainty by quadrature is 0.47 dex for SB\&T galaxies and 0.28 dex for NIS galaxies.

 The slopes are in agreement, therefore both samples are in the main sequence of star formation, although the 
SB\&T sample presents more scatter in that relation. The slopes (Eqs. \ref{eq_intsbt} and \ref{eq_intspi}) are in the range 
of observed values 0.6-1 \citep{2011ApJ...739L..40R}. 


\subsection{Resolved star formation main sequence}

The results are based on 879 clumps from the SB\&T galaxies, and 541 clumps from the NIS galaxies.


\begin{figure*}
\plotone{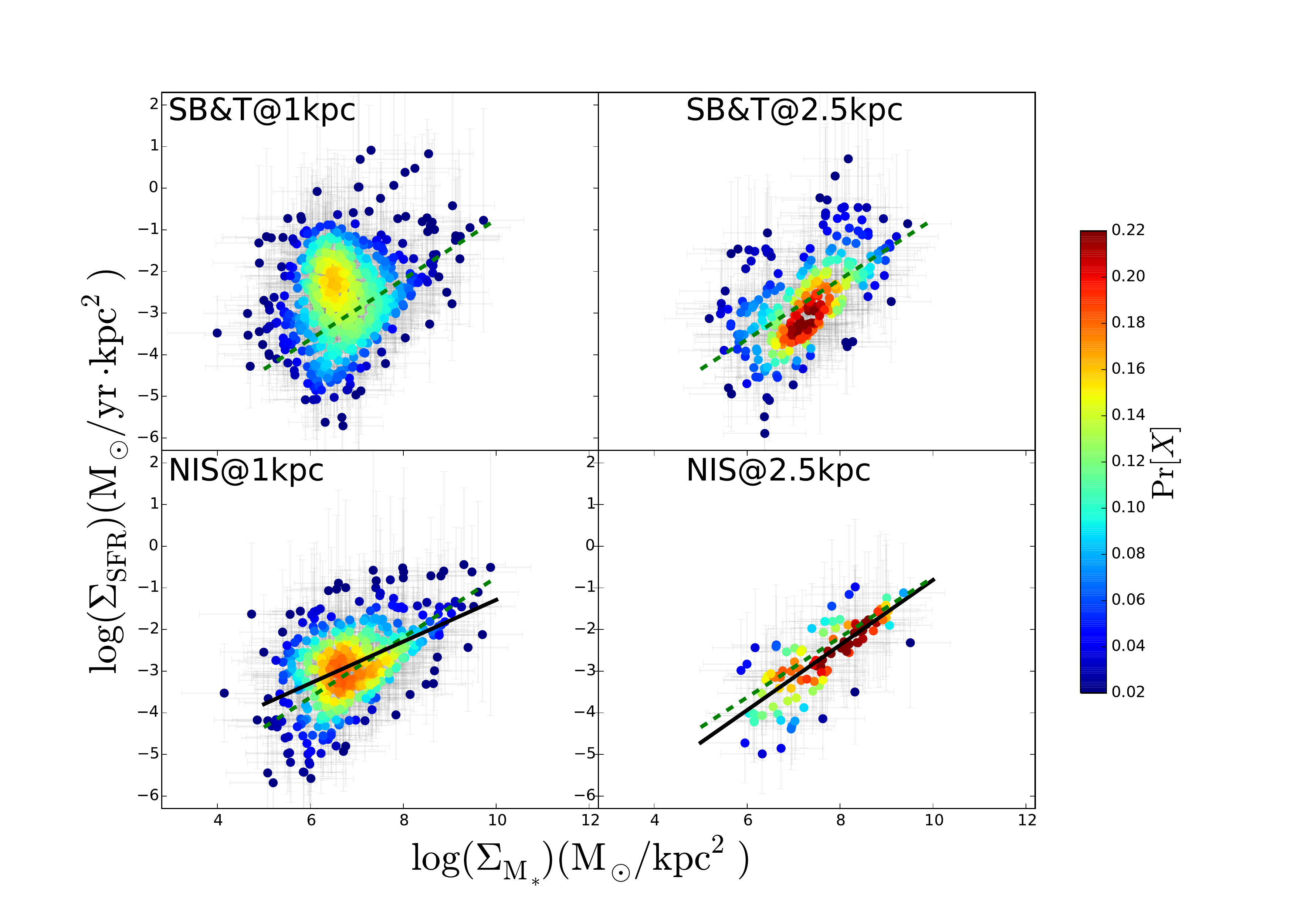}
\caption{Resolved star formation rate per area, $\Sigma_{\rm{SFR}}$, versus resolved stellar mass per area, $\Sigma_{\rm{M_{*}}}$, 
for clumps identified in the SB\&T galaxies (top) and NIS galaxies 
(bottom). The solid black lines are the linear fits with variable $x-bin$ size. 
The dashed green line is the resolved star formation main sequence from \citet{2016ApJ...821L..26C}.
\label{fig_mainseq}}
\end{figure*}

In \citet{2016AJ....151...63S} we already showed that the clumps in the SB\&T galaxies have higher SFRs compared to those in NIS. Here, we explore the differences in the SFR between 
SB\&T and NIS galaxies relative to the stellar mass of the clumps. 

We show in Fig. \ref{fig_mainseq} the resolved SFR per area, $\Sigma_{\rm{SFR}}$, versus the resolved stellar mass per area, 
$\Sigma_{\rm{M_{*}}}$.  
\citet{2016ApJ...821L..26C,2017MNRAS.466.1192M,2017MNRAS.469.2806A} have already shown that the resolved star formation main sequence 
holds on kiloparsec scales in nearby galaxies.
 Here, we show that for the 
clumps in the SB\&T galaxies, the resolved star formation main sequence presents a different pattern compared to the clumps 
in the control sample of NIS on 1kpc scales. More precisely, there is no linear correlation between SFR and stellar mass, with a 
large fraction of clumps displaying excess SFR at
$\log(\Sigma_{\rm{M_{*}}})(\rm{M_{\odot}/kpc^2})\sim6.5$. Although a comparable
 cloud of points is seen in the clumps of the NIS galaxies sample on 1kpc scales, it is seen to be weaker than that in the  SB\&T sample. It is notable that when the results are considered on 2.5kpc 
scales, the cloud of points with an SFR excess vanishes in the SB\&T galaxies and also in the NIS. Thus, the resolved star formation main sequence 
does not hold on kiloparsec scales. 

In order to quantify deviations and enhancements compared with the star formation main sequence, we perform 
a variable $x-bin$ size fit to the $\Sigma_{\rm{SFR}}$-$\Sigma_{\rm{M_{*}}}$ data points for NIS on 1 and 2.5kpc scales.
The variable $x-bin$ size fit allow us to weight by the density of data points, assuming a constant number of data 
points in each bin. We know that the resolved star formation main sequence for NIS on 1 kpc scales deviates from a 
linear relation (Fig. \ref{fig_mainseq} bottom-left). Thus, the linear fit in this case is an approximation 
to measure the deviation of the SB\&T clumps from the NIS clumps on 1 kpc scales.
The results of the linear fits for the NIS galaxies are:
 
\begin{equation}
\begin{array}{ll}
\log(\Sigma_{\rm{SFR}})\left(\frac{\rm{M_{\odot}}}{\rm{yr\thinspace kpc^2}}\right)=\\
(0.50\pm0.06)\log(\Sigma_{\rm{M_{*}}})\left(\frac{\rm{M_{\odot}}}{\rm{kpc^2}}\right)-(6.3\pm0.4),
 \label{eq_mseq1}
 \end{array}
\end{equation}

for $1\thinspace \rm{kpc}$ scales, and

\begin{equation}
\begin{array}{ll}
 \log(\Sigma_{\rm{SFR}})\left(\frac{\rm{M_{\odot}}}{\rm{yr\thinspace kpc^2}}\right)=\\
 (0.78\pm0.05)\log(\Sigma_{\rm{M_{*}}})\left(\frac{\rm{M_{\odot}}}{\rm{kpc^2}}\right)-(8.6\pm0.4),
 \label{eq_mseq25}
\end{array}
 \end{equation}

for $2.5\thinspace \rm{kpc}$ scales. The scatter of the resolved main sequence of star formation in NIS after removing 
the mean uncertainty of the estimated SFR by quadrature is 
$0.41\thinspace \rm{dex}$ for $1\thinspace \rm{kpc}$ scales, and $0.36\thinspace \rm{dex}$ for $2.5\thinspace \rm{kpc}$ scales, 
both of these values are larger than those found by \citet{2016ApJ...821L..26C},  although similar to those found by \citet{2017MNRAS.466.1192M,2017MNRAS.469.2806A},  and larger compared 
to the scatter of the integrated main sequence of star formation for NIS galaxies.
These results, shown as a solid black line in Fig. \ref{fig_mainseq} (bottom), 
show that the resolved star formation main sequence for the two sets of galaxies is different on 1kpc scales. 
The slope for the NIS is lower on 1kpc scales due to the excess of SFR at 
$\log(\Sigma_{\rm{M_{*}}})(\rm{M_{\odot}/kpc^2})\sim6.5$, which is also present in the SB\&T clumps on those scales.

\begin{figure}
\plotone{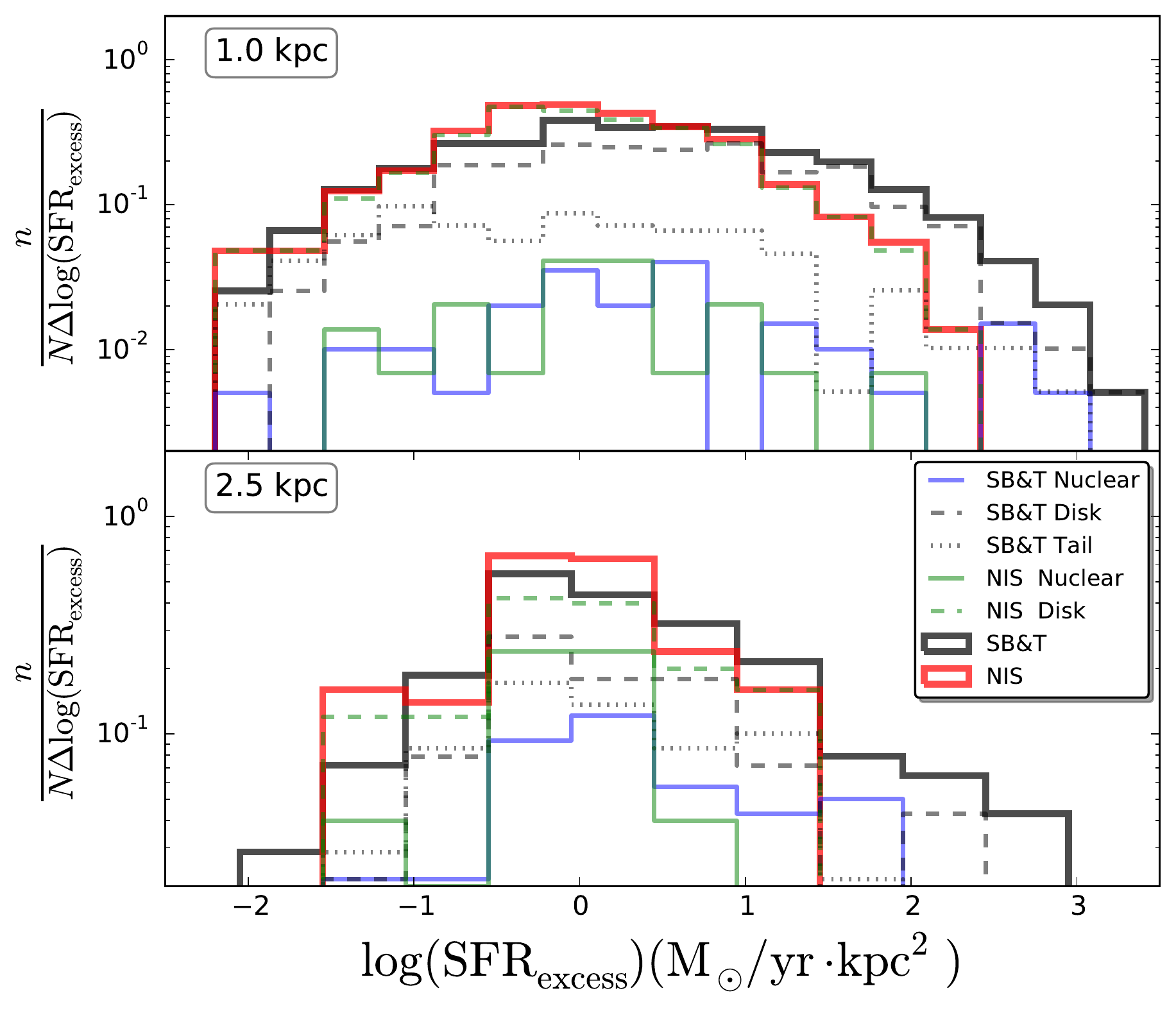}
\caption{Histograms of the SFR excess of the clumps in the SB\&T galaxies (solid black line), and in NIS galaxies (solid red line), normalized 
to the total number of clumps in the SB\&T galaxies and NIS, respectively. 
The histograms of the SFR excess of the clumps in tails (black dotted line), in disks (dashed black line), and in the nuclei (blue solid line) of the 
SB\&T galaxies, are normalized to the total number of clumps in the SB\&T galaxies. 
The histogram of SFR excess of the clumps in disks (green dashed line) and in the nucleus (green solid line) of the NIS galaxies, is normalized to the total number of 
clumps in NIS galaxies.
Top: Clumps on $1\rm{kpc}$ scales. Bottom: Clumps on $2.5\rm{kpc}$ scales. The SFR excess is defined 
as the difference between the SFR derived from the SED modeling and the one derived from Eqs. \ref{eq_mseq1} and \ref{eq_mseq25}.
\label{fig_sfrexcess}} 
\end{figure}

We define the $\rm{SFR_{excess}}$ as the difference between the SFR surface density obtained by the SED modeling and the SFR obtained using Eqs. \ref{eq_mseq1} and \ref{eq_mseq25}, 
and using the stellar mass 
from Tab. \ref{tab_cig_out}. 
 The $\rm{SFR_{excess}}$ represents the deviation of the observed SFR from that expected, as derived from the 
resolved main sequence of star formation determined for NIS galaxies.
We show in Fig. \ref{fig_sfrexcess} the histograms of the $\rm{SFR_{excess}}$ normalized to the total number of clumps in the SB\&T galaxies (solid black line), 
and the number of clumps in NIS (solid red line). We also show the histograms of the SFR excess for the clumps in tails (black dotted line),  
in the disks (dashed black line), and in the nucleus (blue solid line), of the SB\&T galaxies, normalized to the total number of clumps in the SB\&T galaxies. We observe that 
there is a population of clumps with higher SFR excess in the SB\&T galaxies, present in the tail, disk, and nuclear clumps,
compared to the clumps 
in NIS on both $1\rm{kpc}$ scales (top), and  $2.5\rm{kpc}$ scales (bottom). SFR excesses in the clumps in 
the SB\&T galaxies are probably induced 
by the interaction, and make a very good case for studying the triggered star formation regime in galaxy pairs. In 
the higher SFR excess clump population, the star formation is not a local process as claimed 
by  \citet{2016ApJ...821L..26C}, but a global process, because it is affected and enhanced by the interaction. 

On 1kpc scales, the resolved 
star formation main sequence is different compared to that at $2.5\rm{kpc}$ even in NIS galaxies, pointing toward a break of the star formation main sequence on smaller 
scales. 
 \citet{2016ApJ...821L..26C,2017MNRAS.466.1192M,2017MNRAS.469.2806A} did not observe this break probably because they are based on 
pixel-to-pixel SED fitting, while in this work we perform the SED fitting based on clumps, {\it i. e. }, in maximum peaks of 
star formation, and we subtracted the local galaxian background for each clump. Thus, finding an excess of SFR with respect to the 
stellar mass is more plausible with our method.

\begin{figure*}
\gridline{\fig{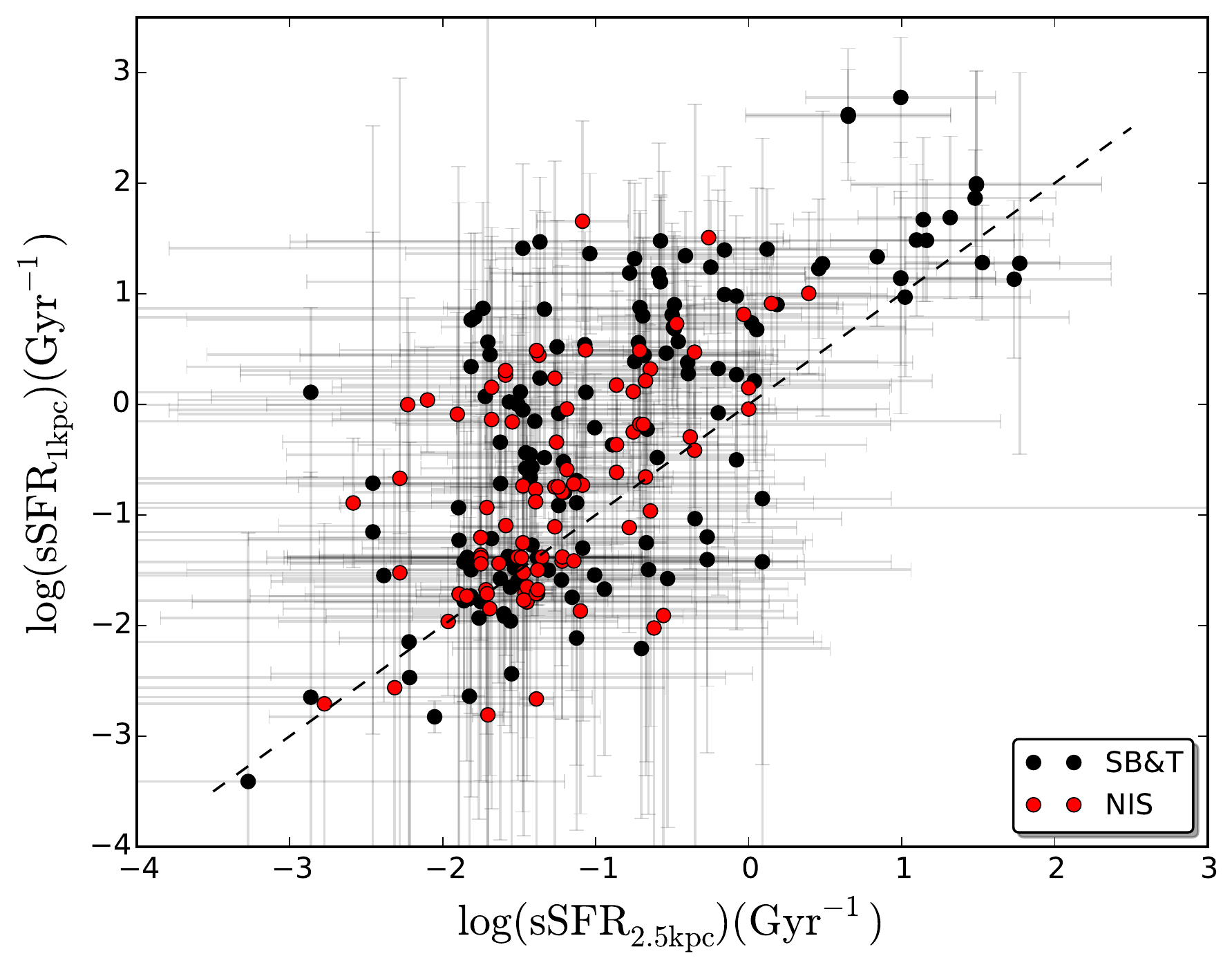}{0.5\textwidth}{(a)}
          \fig{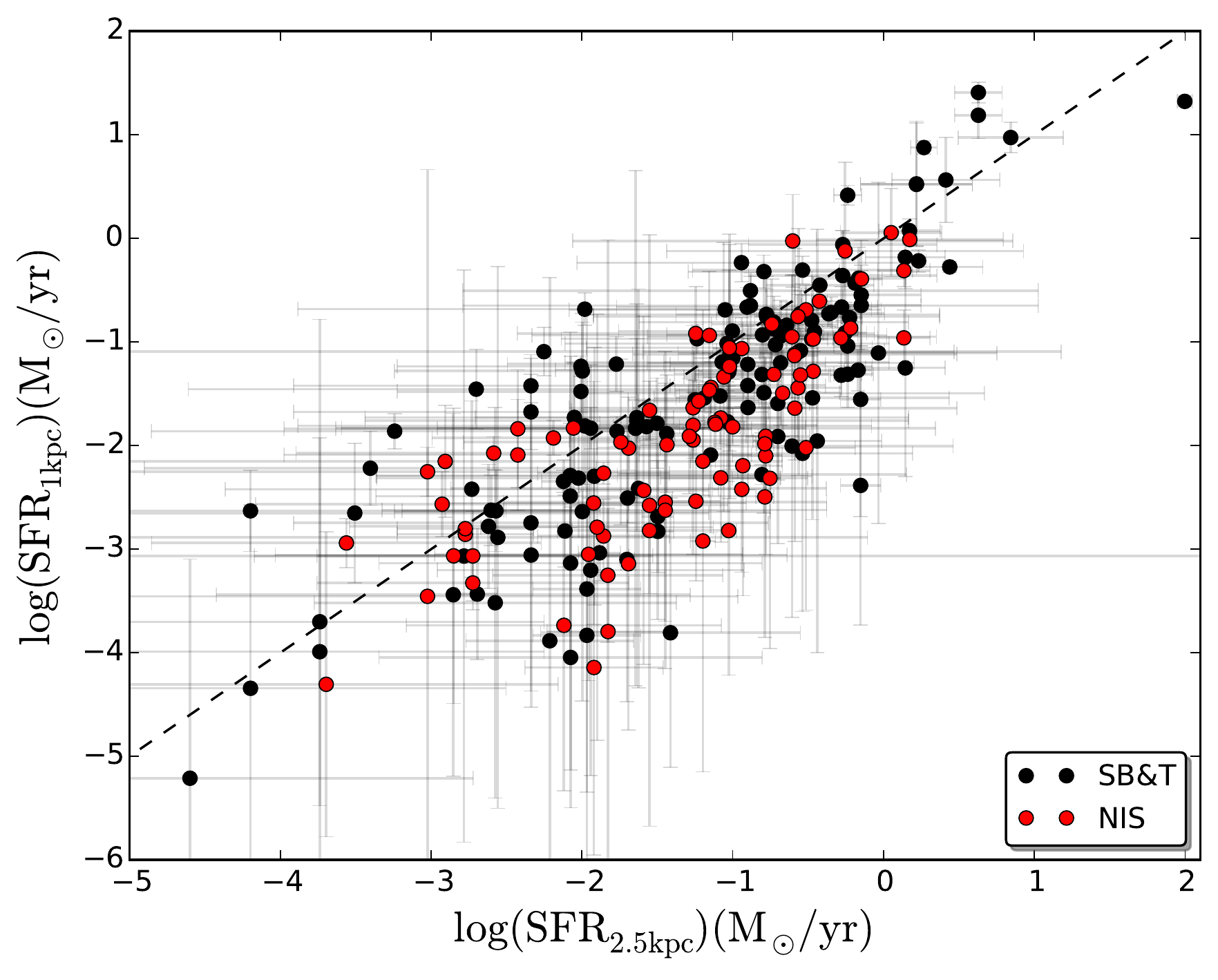}{0.5\textwidth}{(b)}
          }
\gridline{\fig{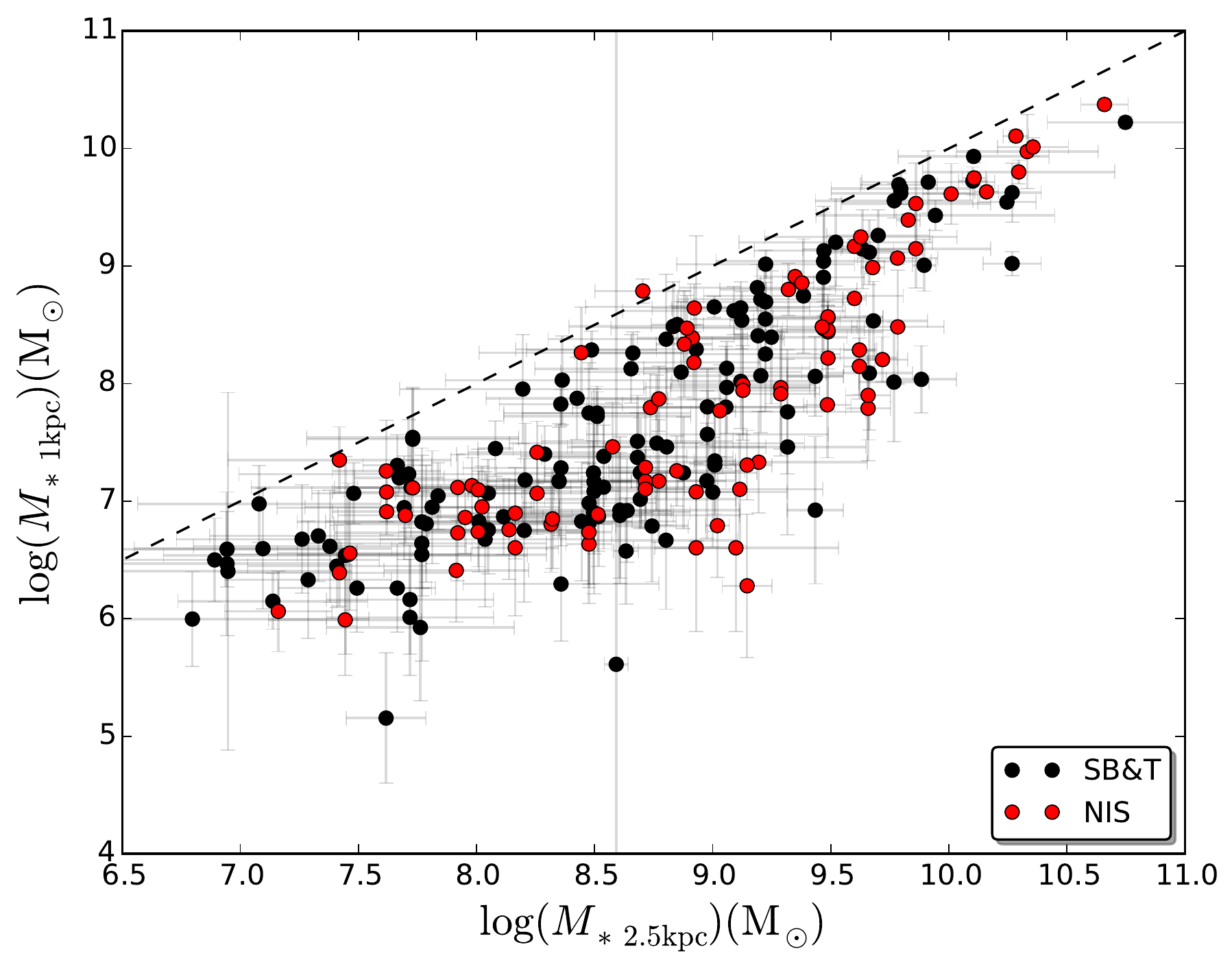}{0.5\textwidth}{(c)}
          \fig{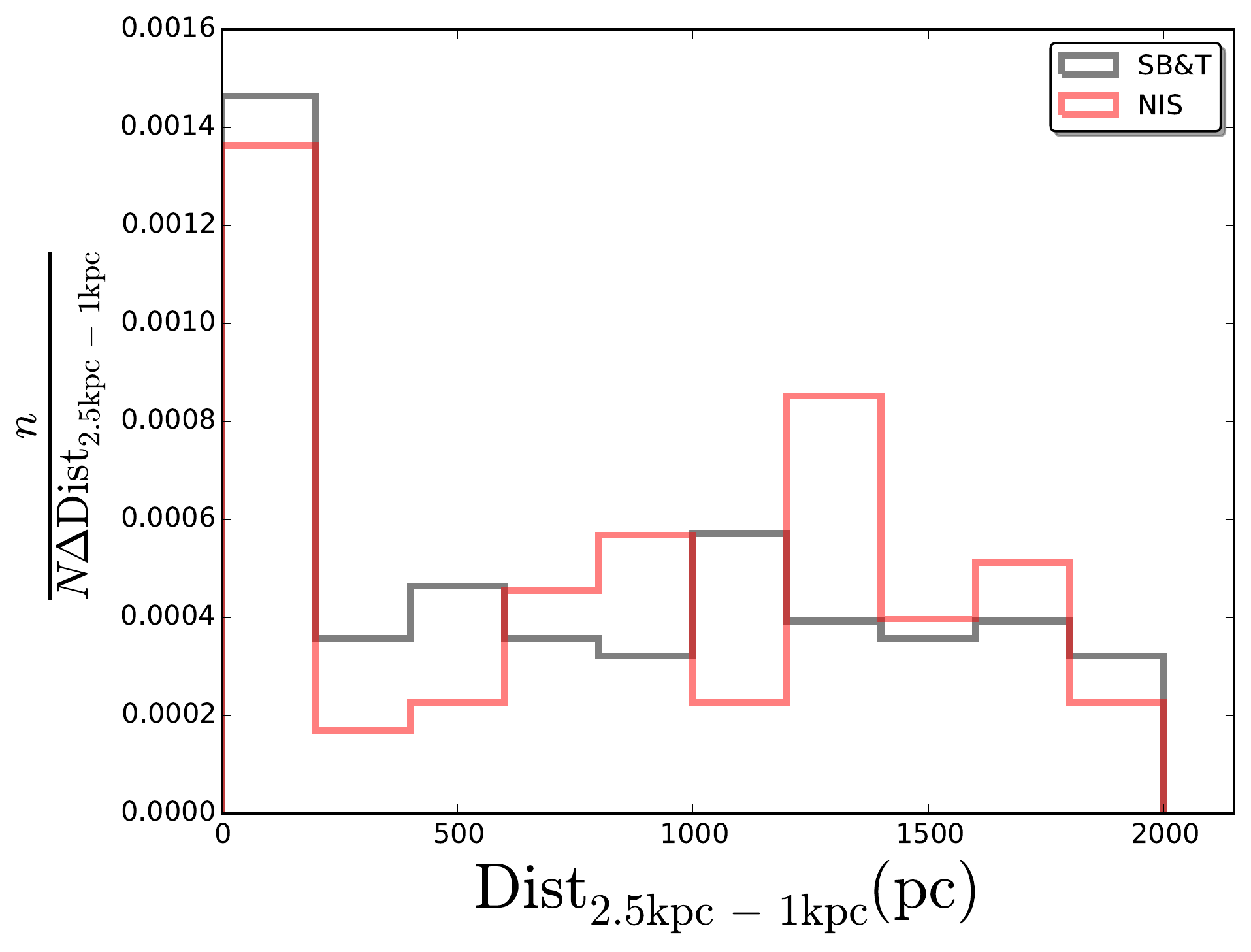}{0.5\textwidth}{(d)}
          }

\caption{
(a) Specific SFR at 1kpc scales, $\rm{sSFR_{1kpc}}$, versus specific SFR at 2.5kpc scales, $\rm{sSFR_{2.5kpc}}$.
(b) SFR at 1kpc scales, $\rm{SFR_{1kpc}}$, versus SFR at 2.5kpc scales, $\rm{SFR_{2.5kpc}}$. 
(c) Stellar mass at 1kpc scales, $M_{\rm{*\, 1kpc}}$, versus stellar mass at 2.5kpc scales, $M_{\rm{*\, 2.5kpc}}$.
(d) Histograms of the distance between center of clumps at 2.5kpc scales and clumps at 1kpc scales, $\rm{Dist_{2.5kpc\, - \, 1kpc}}$.
All of the plots 
are for those clumps 
at 2.5kpc scales which have 1kpc clumps inside them.
Clumps from the SB\&T sample are in black, and clumps from the NIS sample are in red. 
The dashed lines are the 1 to 1 relation. 
\label{fig_ssfr_1_25}}
\end{figure*}

%

We have cross-correlated the two sets of clumps (1kpc and 2.5kpc) to 
find 1kpc clumps within 2.5kpc clumps.
In Fig. \ref{fig_ssfr_1_25} we show the specific SFR (a), the SFR (b), and the stellar mass (c), at both scales for clumps at 2.5kpc, which have one or more 1kpc clump inside them. 
The specific SFR at 1kpc scales is larger compared to that of clumps at 2.5kpc scales, which explains the larger SFR excess found at 1kpc 
scales for both SB\&T (black circles) and NIS (red circles) samples. On average, the sSFR is 4 times larger at 1kpc scales compared to 2.5kpc scales. 
This is due to the fact 
that the SFR is more centrally concentrated than the older stellar population as can be seen in Figs. \ref{fig_ssfr_1_25} (b) and (c), since the SFR 
at 1kpc vs SFR at 2.5kpc distribution is closer 
to the one to one relation at both scales, while the stellar mass at 2.5 kpc is larger than the stellar mass at 1kpc scales.
In addition, in Fig. \ref{fig_ssfr_1_25} (d) we plot the histograms of the distances between the centers of the clumps at 2.5kpc and at 1kpc, $\rm{Dist_{2.5kpc\, - \, 1kpc}}$, 
for those 2.5kpc clumps which have one or more 1kpc inside them. The distances are dominated by a population of clumps at both scales having the same centers, which means 
that the strongest star formation tends to occurs at the center of large old stellar clumps. However, there is a population of clumps at both scales having very different centers 
($\rm{Dist_{2.5kpc\, - \, 1kpc}}>1\rm{kpc}$). 
If we consider that 1 kpc clump is within 
a 2.5kpc clump if at least half of it is completely inside, just 47 clumps at 2.5kpc have two or more 1kpc clumps within them out of 381 2.5kpc identified clumps. 
Then, we can neglect the effect of blending.


\subsection{Recently induced star formation}

\begin{figure*}
\plotone{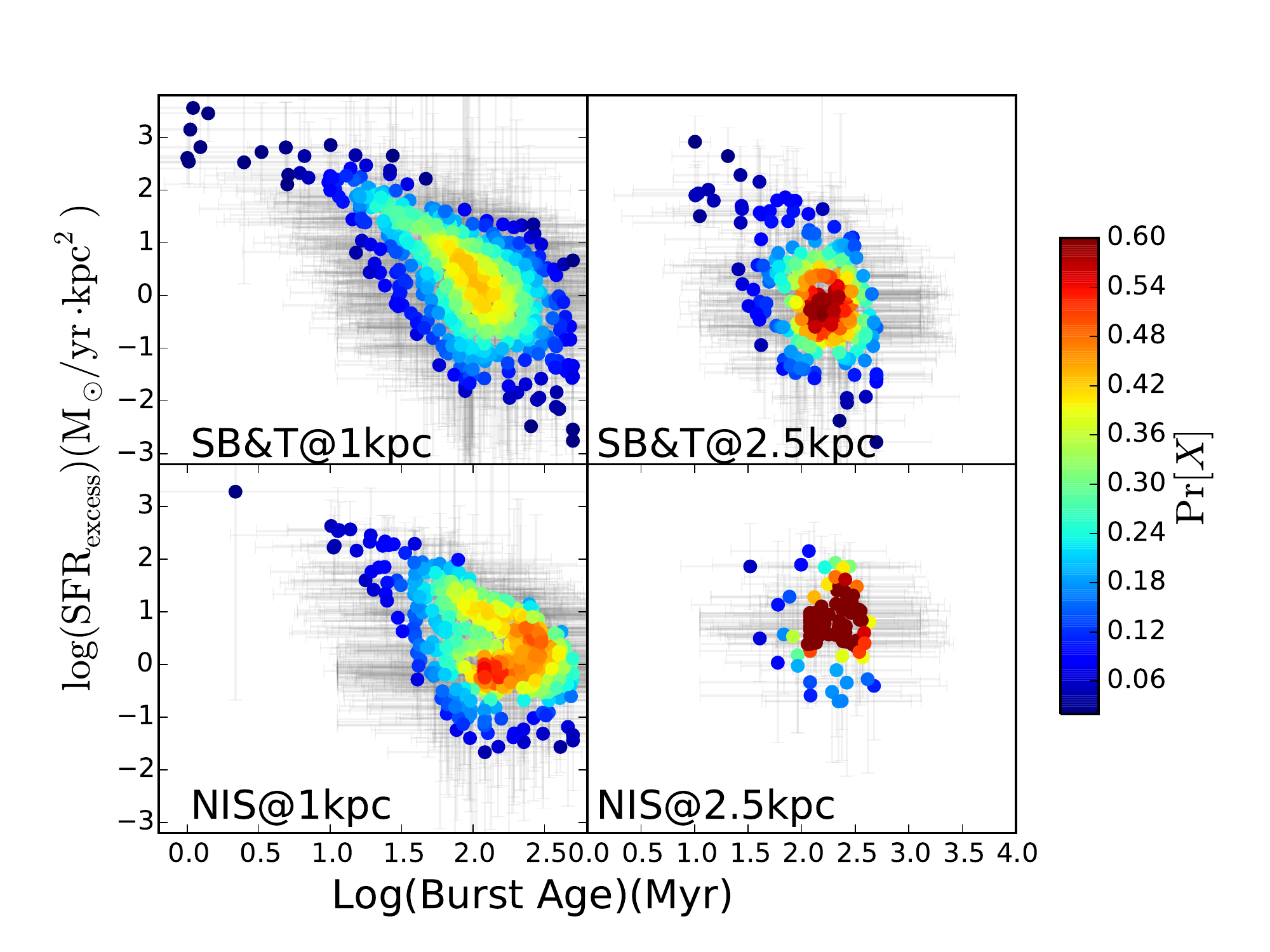}
\caption{SFR excess versus the age of the recent starburst, color coded with the  PDF. 
The figure displays clumps in the SB\&T galaxies (top) and in 
the NIS galaxy sample (bottom). The SFR excess is defined 
as the difference between the SFR derived from the SED modeling and that derived from Eqs. \ref{eq_mseq1} and \ref{eq_mseq25}.
\label{fig_age_sfr}}
\end{figure*}

In order to explore the possible connection between the higher SFR excess of clumps in the SB\&T galaxies and their recent interaction history, 
we compared the derived age of the recent burst ($t_2$ in Eq. \ref{eq_sfh}) from the SED fitting with 
the SFR excess. 
We plot in Fig. \ref{fig_age_sfr} the SFR excess versus the age of the recent burst for clumps in the SB\&T galaxies (top) and in the NIS galaxies (bottom), 
color coded with the  PDF. 
Fig. \ref{fig_age_sfr} shows that the SFR excess depends on the age of the recent burst of star formation; the younger 
the recent burst, the higher the SFR excess. Additionally, the density of data points shows that 
the SB\&T galaxies have a population of clumps which 
have a younger recent burst of star formation, notably at $\log(\rm{Burst\thinspace Age})\sim1.9$, and also at 
$\log(\rm{Burst\thinspace Age})<1$, compared to the NIS galaxies. Therefore, the triggering of the SFR excess in the SB\&T galaxies is evidently 
due to a recent event such as the interaction with a companion galaxy. 

\begin{figure}
\plotone{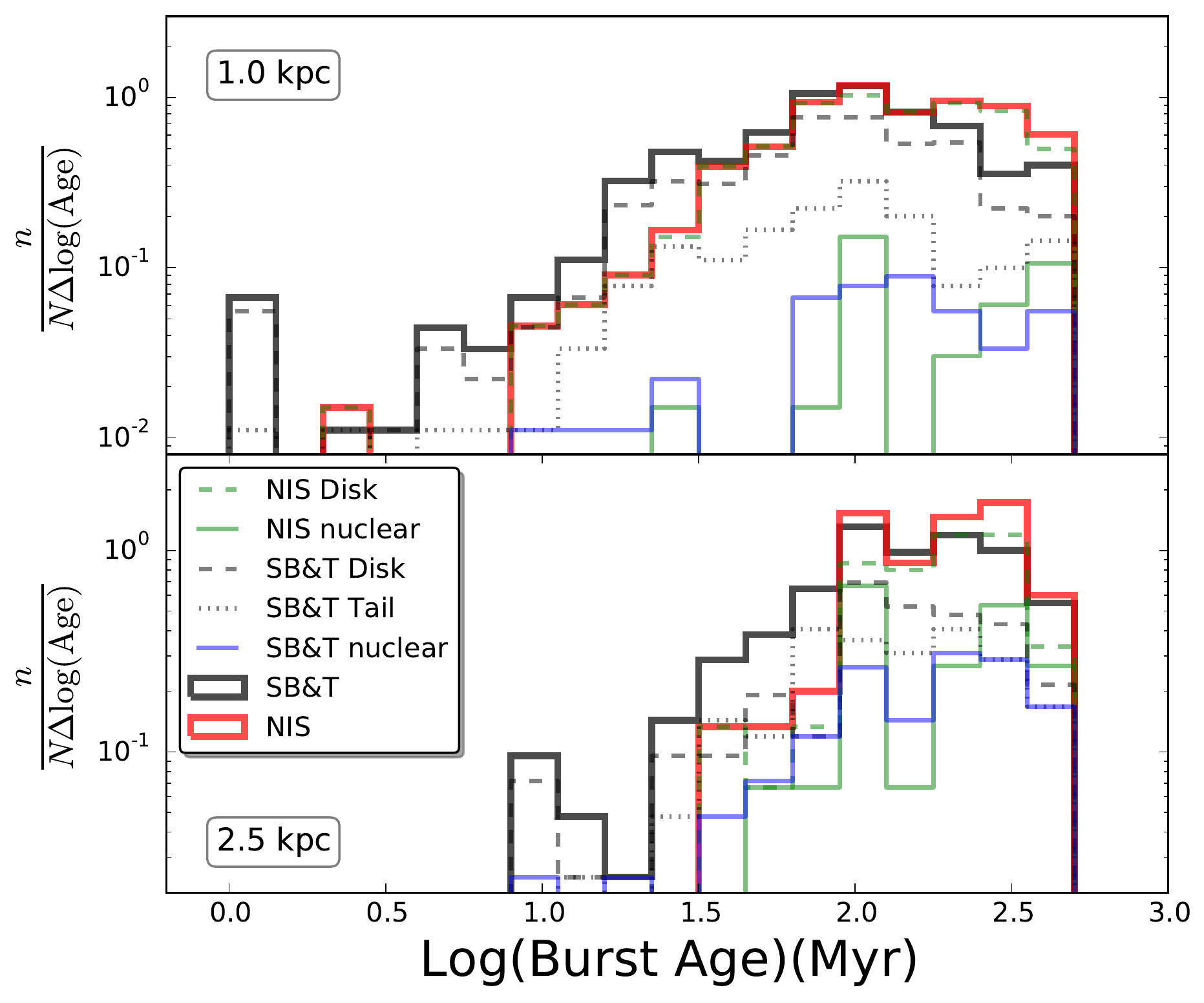}
\caption{Age of the recent starburst histograms for clumps in the SB\&T galaxies (solid black line), and 
for clumps in the NIS galaxies (solid red line), normalized by the total numbers of clumps in the 
SB\&T galaxies and the total number of clumps in the NIS galaxies, respectively. Top: Clumps 
at 1kpc scales. Bottom: Clumps at 2.5kpc scales. The black dotted lines are the age of the recent starburst histograms
of the clumps in tails, the dashed black lines are the age of the recent starburst histograms of the clumps 
in the disks, and the solid blue lines are the recent starburst histograms of the clumps 
in the nucleus for clumps in the SB\&T sample, and normalized by the total number of clumps in the SB\&T sample.
The solid and dashed green lines are the age of the recent starburst histograms of the clumps 
in the nucleus and in the disks, respectively, for clumps in the NIS galaxies, 
and normalized by the total number of clumps in the NIS galaxies.
\label{fig_ageburst}}
\end{figure}

Histograms of the age of the recent burst (Fig. \ref{fig_ageburst}) show that there is a population of clumps in the SB\&T galaxies (solid black lines) with smaller 
ages compared to the NIS galaxies (solid red line) on both scales. 
Younger recent burst ages are found in the tails (black dotted line), the disks (dashed black line), and the nuclei (blue solid line) of the SB\&T galaxies. 
These results show that there are more recent bursts of star formation in the clumps of the SB\&T galaxies induced by the interactions, which 
enhance the observed SFR excess. 

\section{SFR radial profile}

\begin{figure}
\plotone{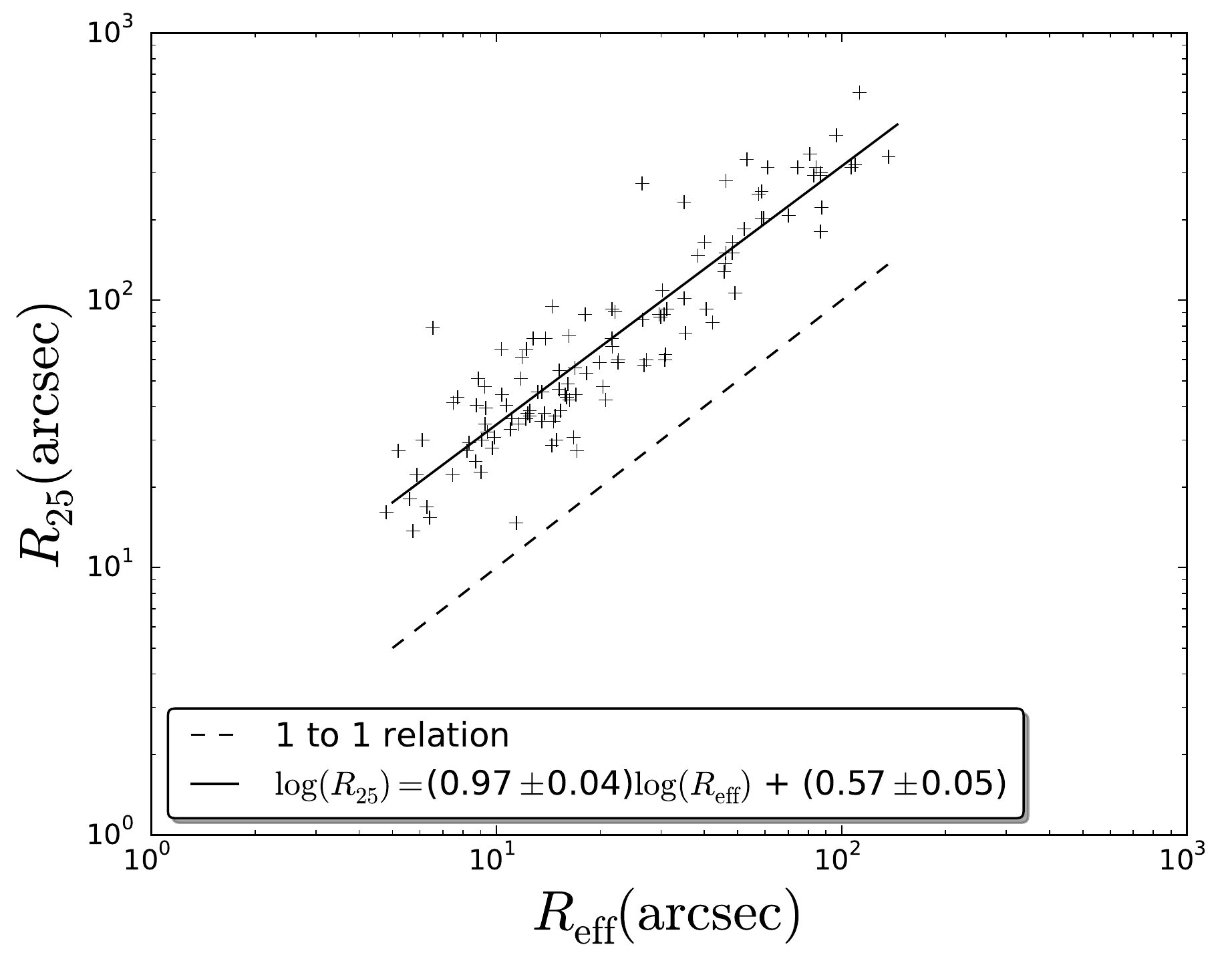}
\caption{Isophotal radius at $25\thinspace\rm{mag/arcsec^2}$ in the B-band, $R_{25}$, versus the effective radius in the J band 
from 2MASS, $R_{\rm{eff}}$, for SB\&T and NIS galaxies. The solid line is the linear fit to the data points, while the dashed line 
is the 1 to 1 one relation.
\label{fig_rad25_rads}
}
\end{figure}

\begin{figure*}
\plotone{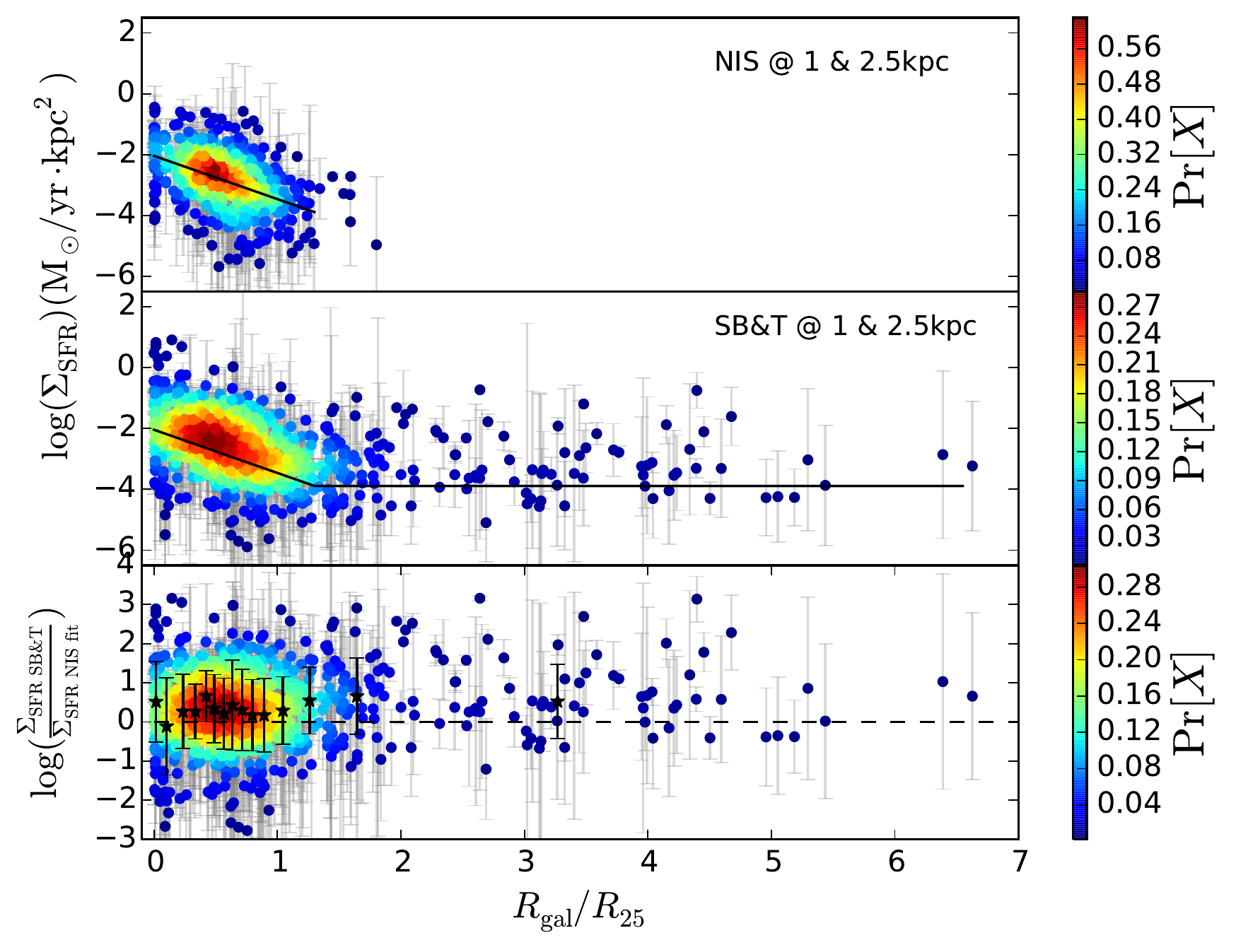}
\caption{Top: SFR surface density, $\Sigma_{\rm{SFR}}$, versus the galactocentric radius normalized by the 
isophotal radius at $25\thinspace\rm{mag/arcsec^2}$ in the B-band, $R_{\rm{gal}}/R_{25}$, of the clumps 
in disks from the NIS galaxies, color coded with the  PDF. 
The solid black line is the variable $x$-bin size 
fit to the points.
Middle: SFR surface density, $\Sigma_{\rm{SFR}}$, versus the galactocentric radius normalized by the 
isophotal radius at $25\thinspace\rm{mag/arcsec^2}$ in the B-band, $R_{\rm{gal}}/R_{25}$, of the clumps 
in disks and tails from the SB\&T galaxies sample, color coded with the  PDF. 
The solid black line is the fit to the 
points of the clumps in NIS galaxies (top panel), and for $R_{\rm{gal}}/R_{25}>1.15$ we extrapolate the value of the fit as a 
constant one. 
Bottom: Ratio between the SFR surface density for clumps in the SB\&T galaxies and the SFR surface density derived from 
the linear fit for clumps in NIS (Eq. \ref{eq_sfrrad_spi}), 
$\Sigma_{\rm{SFR\thinspace SB\&T}}/\Sigma_{\rm{SFR \thinspace NIS \thinspace fit}}$, 
versus the galactocentric radius normalized by the 
isophotal radius at $25\thinspace\rm{mag/arcsec^2}$ in the B-band, $R_{\rm{gal}}/R_{25}$, of the clumps 
in disks and tails from the SB\&T galaxies sample, color coded with the  PDF. 
We show the 
median values of the $\Sigma_{\rm{SFR\thinspace SB\&T}}/\Sigma_{\rm{SFR \thinspace NIS \thinspace fit}}$ 
for a variable $x$-bin size as a black star symbols,  with their standard deviations as error bars. 
\label{figsfrrad}
}
\end{figure*}

The SB\&T sample of galaxies is composed of galaxy pairs in an early-intermediate stage of the merger process, 
while advanced mergers are excluded. Therefore, the distortions are small enough to  be able to study 
the SFR radial profile for the clumps in the SB\&T sample, in order to see the 
radial variation of the SFR enhancement.

We normalized the galactocentric radius of 
each identified clump by the isophotal radius at $25\thinspace\rm{mag/arcsec^2}$ in the B-band, in order to 
compare all the galaxies from both samples together.
We obtained the inclinations, position angles, and lengths of the major axis at the isophotal level 
$25\thinspace\rm{mag/arcsec^2}$ in the B-band, from the Hyperleda database \citep{2014A&A...570A..13M}\footnote{http://leda.univ-lyon1.fr/} for each galaxy 
(see Tabs. \ref{tab_sbtsample} and \ref{tab_nissample}). 
 We show in Fig. \ref{fig_rad25_rads} the 
isophotal radius at $25\thinspace\rm{mag/arcsec^2}$ in the B-band, $R_{25}$, versus the effective radius in the J band 
from 2MASS, $R_{\rm{eff}}$, for SB\&T and NIS galaxies. 
We obtained J band $R_{\rm{eff}}$  from the 2MASS All-Sky Extended Source Catalog \citep{2006AJ....131.1163S}. 
Since the surface brightness is independent of distance, the choice of 
isophotal or effective radius is just affected by a constant factor, thus the selection of the isophotal radius does 
not affect the results presented below.

For the SB\&T galaxies, we obtained those parameters for each individual galaxy and 
associate each clump with one of the galaxies  to normalize the galactocentric radius of each clump with the corresponding isophotal radius of his galaxy. 
The Galaxy column in Tab. \ref{tableclumps_photometry} refers to the specific galaxy from the galaxy pair the clump is associated with.

 Several studies show that the spatial distribution of the SFR in spirals approximately follows an exponential profile 
\citep{1983AJ.....88..296H,1993A&AS..102..229A,1994ApJ...430..142R,2006AJ....131..716K}. Thus, 
we plot in Fig. \ref{figsfrrad} (top) the SFR surface density, $\Sigma_{\rm{SFR}}$, of the clumps 
in disks from the NIS galaxies, versus the galactocentric radius normalized by the 
isophotal radius at $25\thinspace\rm{mag/arcsec^2}$ in the B-band, $R_{\rm{gal}}/R_{25}$, color coded with the  PDF, and 
we perform a variable $x$-bin size fit to the data, obtaining 

\begin{equation}
 \log(\Sigma_{\rm{SFR}})=(-1.42\pm0.10)\thinspace R_{\rm{gal}}/R_{25}-(2.04\pm0.06).
 \label{eq_sfrrad_spi}
\end{equation}

The variable $x-bin$ size fit allows us to weight by the density of data points, assuming a constant number of data 
points in each bin. 

In the middle panel of Fig. \ref{figsfrrad} we plot $\Sigma_{\rm{SFR}}$ versus  $R_{\rm{gal}}/R_{25}$, of the clumps 
in the disks and tails of the SB\&T galaxies, color coded with the PDF. We add to this plot 
the fit from the clumps in NIS (top panel) to compare how the SFR radial profile differs in both samples. 
 We extrapolate the last value of Eq. \ref{eq_sfrrad_spi} for the last radial bin for the NIS galaxies to larger radii, 
to compare with the tidal features of the SB\&T galaxies. 

 $\Sigma_{\rm{SFR}}$ is on average larger in SB\&T clumps compared to NIS clumps.
 To study in more detail the differences between the SFR radial profiles in the SB\&T and NIS 
galaxies samples, we derive $\Sigma_{\rm{SFR\thinspace SB\&T}}/\Sigma_{\rm{SFR \thinspace NIS \thinspace fit}}$, which is 
the ratio between the observed $\Sigma_{\rm{SFR}}$ and that derived from 
Eq. \ref{eq_sfrrad_spi} and the extrapolation using the corresponding $R_{\rm{gal}}/R_{25}$ value for the clumps in 
disks and tails from the SB\&T galaxies. We plot 
$\Sigma_{\rm{SFR\thinspace SB\&T}}/\Sigma_{\rm{SFR \thinspace NIS \thinspace fit}}$ versus 
$R_{\rm{gal}}/R_{25}$ in the bottom panel of Fig. \ref{figsfrrad}, where we show how the 
SFR surface density increases toward the central parts of the SB\&T galaxies 
compared to the NIS between $R_{\rm{gal}}/R_{25}\in[0.4,0.9]$, which is 
in agreement with theoretical models of galaxy interactions, where gas inflows are 
produced by the loss of axisymmetry.  There is less SFR enhancement toward inner regions 
$R_{\rm{gal}}/R_{25}<0.4$, except the nuclear regions, which present a median enhancement of 2.4.

In Fig. \ref{figsfrrad} (bottom) we also show that the SFR surface density increases toward
the external parts of the SB\&T galaxies compared to the NIS between $R_{\rm{gal}}/R_{25}\in[1,6.5]$. 
 We extrapolate the exponential fit derived (Eq. \ref{eq_sfrrad_spi}) to external regions using the value of the last radial bin from 
 the NIS galaxies, since we do not 
observe clumps in the NIS galaxies beyond $R_{\rm{gal}}/R_{25}\sim2$. Although this extrapolation may not represent the real values, 
 it is a conservative upper limit of $\Sigma_{\rm{SFR}}$ for the clumps in NIS.
The SFR enhancement in the external parts of galaxy mergers is highly debated because the evidence of the enhancement 
has been based on individual cases. Here, we present evidence for a larger sample of galaxy pairs in 
intermediate-early stages of interaction, where the SFR is clearly enhanced far from the nucleus.  We obtain an 
SFR enhancement for clumps in the SB\&T galaxies where $R_{\rm{gal}}/R_{25}> 2$ of 
$\frac{\Sigma_{\rm{SFR\thinspace SB\&T}}}{\Sigma_{\rm{SFR \thinspace NIS \thinspace fit}}}>2.1$.

\section{Discussion \& Conclusion}

We present  stellar population synthesis analysis of 879 clumps from the SB\&T galaxy sample, and 541 clumps from the NIS galaxy sample using the CIGALE SED modeling code,
and UV, optical, and IR photometry 
of the clumps. Using CIGALE we 
obtained SFRs, stellar masses, ages of the most recent burst, and fractions of the most recent burst, for the identified clumps.

The resolved star formation main sequence was presented by  \citet{2016ApJ...821L..26C,2017MNRAS.466.1192M,2017MNRAS.469.2806A} for nearby galaxies, 
where they showed that it does hold on kiloparsec scales ($[1-2\rm{kpc}]$). However, we 
find that for the identified clumps at 1kpc scales, the main sequence begins to breakdown in the NIS galaxies, and more intensely in the SB\&T galaxies, while for the 
clumps at 2.5kpc scales the main sequence holds,  although it presents a higher scatter compared to that of the integrated star formation main sequence for NIS galaxies. 
 We selected those scales in an effort to study star formation in higher resolution (1kpc) due to the proximity of the sources to us (those with $D<67\thinspace\rm{Mpc}$), 
and also to study 
star formation for all the galaxies. We were limited by the most distant galaxy, Arp 107, at $142\thinspace\rm{Mpc}$, and the resolution of the GALEX and {\it Spitzer} 24 $\mu$m images, 
which approximately corresponds to $2.5\thinspace\rm{kpc}$ at $142\thinspace\rm{Mpc}$.
We show that 
the resolved star formation main sequence breaks down  at small scales (between $1\thinspace\rm{kpc}$ and $2.5\thinspace\rm{kpc}$). 
As in the case of the Kennicutt-Schmidt law, which  breaks down for sub-kpc scales \citep{2008AJ....136.2846B,2010ApJ...722L.127O}, a break 
is expected a small scales since stellar mass and star formation rate trace different properties of the star formation process, and these breaks could 
be used to constrain unknown quantities related to the star formation such as the duration of different star formation phases \citep{2014MNRAS.439.3239K}. 

The breakdown is more notable in the clumps from the SB\&T galaxies, where the SFR is higher per stellar mass compared to the clumps in NIS galaxies. The 
SFR excess in the SB\&T galaxies is probably triggered from the interactions, since they drive gas flows, increase turbulence, and compress gas. Therefore, 
at least in the nearby universe, the SFR surface density and the stellar mass 
surface density relation was affected by the environment, where galaxy pairs present higher SFR excess. 
 Mergers should not be important drivers of the SFR enhancement observed at higher redshifts \citep{2014ARA&A..52..415M}, 
because the star formation main sequence has been 
observed to be tight even at high redshifts \citep{2011ApJ...739L..40R,2014ApJS..214...15S}. Thus, higher 
gas fractions have been proposed as a mechanism to enhance SFR at higher redshift, and when mergers occur the SFR is already saturated 
\citep{2017MNRAS.465.1934F}.

 We show that the scatter of the integrated star formation main sequence is larger for SB\&T galaxies compared to NIS galaxies.
However, the 
star formation main sequence evolves with redshift, and so the discrimination between the main sequence and the starburst regime could also evolve.
Whether or not mergers drive higher star formation at earlier epochs, the clumps presented here that have an excess in their SFR due to 
higher gas fractions enhanced by gas inflows due to the interaction, and are thus
excellent laboratories to test models of star formation, see e. g. \citet{1997RMxAC...6..165E,1997ApJ...481..703S,2007ApJ...670..237B,2012ApJ...751...77Z,2014ApJ...793...84Z,2017arXiv170600106K}, 
especially in an enhanced regime such as the clumpy star formation observed 
at higher redshifts \citep{2009ApJ...701..306E,2011ApJ...739...45F,2015ApJ...800...39G}.

Evidence in favor of a deviation from the KS law of star formation is the extended SFR excess 
reported here in the external parts of the SB\&T galaxies in comparison with the 
clumps in NIS. Galaxy simulations assuming only a KS law of star formation are unable to predict 
the extended SFR excess in galaxy collisions \citep{2015MNRAS.448.1107M}. 
The classical picture of gas inflows toward the central parts of merging galaxies is not enough to 
explain the extended enhanced star formation.  Collisionally driven waves \citep{1999PhR...321....1S}, tidal tails \citep{2013LNP...861..327D},  
and shock-induced star formation \citep{2004MNRAS.350..798B,2010MNRAS.407...43C} have been proposed 
as mechanisms to induce extended star formation in galaxy collisions. Also, \citet{2011EAS....51..107B,2013MNRAS.434.1028P,2014MNRAS.442L..33R} presented 
simulations with enough resolution to capture the turbulence of the cold gas, which predict deviations from the KS law of star 
formation, showing that compressive modes of turbulence 
are enhanced in galaxy mergers and produce extended star formation, as we observe in the SB\&T galaxies, and in agreement with 
the velocity dispersion enhancement in interacting galaxies reported by \citet{2015MNRAS.451.1307Z}.

\acknowledgments 
The authors thank Sebasti\'an F. S\'anchez, Mariana Cano-D\'iaz, and Curtis Struck for their helpful comments. 
The authors also thank the anonymous referee, whose comments have
led to important improvements upon the original version of the
paper. 
JZC thanks the DGAPA Postdoctoral fellowships program of the National Autonomous University of Mexico (UNAM), and 
the Luc Binette fellowship. 
MR and JZC acknowledge the grants IN103116 by DGAPA-PAPIIT UNAM and 253085 from CONACYT.
BJS acknowledges support from National Science Foundation Extragalactic Astronomy grant 1311935. 
This research benefited from support and resources from the HPC cluster Atocatl at IA-UNAM.
Based on observations made with the William Herschel Telescope operated on the island of La Palma 
by the Isaac Newton Group of Telescopes in the Spanish Observatorio del Roque de los Muchachos of the Instituto de Astrofísica de Canarias.
This research made use of ASTROPY, a community-developed core Python package for Astronomy \citep{2013A&A...558A..33A}, 
and APLpy, an open-source plotting 
package for Python \citep{2012ascl.soft08017R}.
\software{CIGALE \citep{2009A&A...507.1793N}, Astropy \citep{2013A&A...558A..33A}, APLpy \citep{2012ascl.soft08017R}}
\appendix
\section{SB\&T galaxies sample}
{\catcode`\&=11
\gdef\2014AandA...570A..13M{\citep{2014A&A...570A..13M}}}
\startlongtable
\begin{deluxetable}{cccccccc}
{\def\arraystretch{0.7}
{\setlength{\tabcolsep}{0.5pt}

\tabletypesize{\tiny }
\tablecaption{ SB\&T galaxies sample \tablenotemark{\dag}. \label{tab_sbtsample}}
\tablecolumns{8}
\tablehead{\\
\colhead{System}  & \colhead{Dist($\rm{Mpc}$)}\tablenotemark{a} & \colhead{Galaxy} & \colhead{Morph}\tablenotemark{b} & 
\colhead{logD25}\tablenotemark{c}  &  \colhead{logR25}\tablenotemark{d}  &  \colhead{PA($\rm{\deg}$)}\tablenotemark{e}  &   
\colhead{i($\rm{\deg}$)}\tablenotemark{f} \\ 
}
\startdata
 Arp24  &  33.1   &NGC3445 & SABm & 1.15 & 0.04 & 130. & 27.9\\ 
  & & PGC032784 & Sd & 0.90 & 0.51 & 87.5 & 90.0\\ 
 Arp34  &  72.5   &NGC4613 & Sbc & 0.71 & 0.02 & 15. & 18.6\\ 
  & & NGC4614 & S0-a & 1.00 & 0.06 & 151.6 & 33.2\\ 
  & & NGC4615 & Sc & 1.19 & 0.53 & 120.1 & 76.2\\ 
 Arp65  &  72.0   &NGC0090 & SABc & 0.99 & 0.08 & 120.1 & 34.5\\ 
  & & NGC0093 & Sab & 1.12 & 0.25 & 49.8 & 59.3\\ 
 Arp72  &  53.4   &NGC5994 & SBbc & 0.78 & 0.29 & 93.3 & 62.5\\ 
  & & NGC5996 & SBbc & 1.18 & 0.34 & 1.1 & 66.2\\ 
 Arp82  &  59.2   &NGC2535 & Sc & 1.29 & 0.31 & 62.5 & 63.1\\ 
  & & NGC2536 & SBc & 0.87 & 0.21 & 52.7 & 53.3\\ 
 Arp84  &  55.5   &NGC5394 & SBb & 1.42 & 0.39 & 60. & 70.8\\ 
  & & NGC5395 & SABb & 1.40 & 0.33 & 170.9 & 66.1\\ 
 Arp85  &  12.1   &NGC5194 & SABb & 2.14 & 0.07 & 163.0 & 32.6\\ 
  & & NGC5195 & SBa & 1.74 & 0.10 & 79.0 & 40.5\\ 
 Arp86  &  65.9   &NGC7752 & S? & 0.96 & 0.30 & 105.5 & 63.8\\ 
  & & NGC7753 & SABb & 1.30 & 0.57 & 61.1 & 82.1\\ 
 Arp87  &  104.6   &NGC3808 & SABc & 0.97 & 0.11 & 16.5 & 40.1\\ 
  & & NGC3808B & SBc & 0.87 & 0.35 & 46.1 & 65.5\\ 
 Arp89  &  31.8   &NGC2648 & Sa & 1.50 & 0.47 & 151.0 & 83.7\\ 
  & & PGC024469 & Sbc & 1.00 & 0.58 & 102.3 & 81.4\\ 
 Arp91  &  34.3   &NGC5953 & S0-a & 1.17 & 0.12 & 50. & 43.9\\ 
  & & NGC5954 & SABc & 1.01 & 0.32 & 19.2 & 63.6\\ 
 Arp102  &  104.7   &PGC060067 & E & 0.91 & 0.05 & ... & 35.0\\ 
  & & UGC10814 & SABb & 1.08 & 0.40 & 170.3 & 71.5\\ 
 Arp104  &  50.6   &NGC5216 & E & 1.23 & 0.19 & 54.0 & 83.9\\ 
  & & NGC5218 & SBb & 1.26 & 0.18 & 93.4 & 51.1\\ 
 Arp105  &  126.2   &NGC3561 & S0-a & 1.23 & 0.00 & 175. & 0.0\\ 
  & & UGC06224 & ... & 1.07 & 0.23 & 160.7 & 54.9\\ 
  & & PGC033992 & S0-a & 0.46 & 0.01 & ... & 12.2\\ 
 Arp107  &  141.8   &PGC032620 & SBab & 1.20 & 0.11 & 19.7 & 41.8\\ 
  & & PGC032628 & E & 1.00 & 0.08 & 98.3 & 44.8\\ 
 Arp120  &  14.0   &NGC4435 & S0 & 1.48 & 0.15 & 12.2 & 53.6\\ 
  & & NGC4438 & Sa & 1.96 & 0.36 & 27.0 & 73.2\\ 
 Arp178  &  82.5   &NGC5613 & S0-a & 0.75 & 0.18 & 29.5 & 55.5\\ 
  & & NGC5614 & Sab & 1.38 & 0.08 & 150.0 & 36.2\\ 
  & & NGC5615 & ... & 0.35 & 0.00 & 161.9 & 0.0\\ 
 Arp181  &  132.0   &NGC3212 & SBb & 0.92 & 0.05 & 88.3 & 27.1\\ 
  & & NGC3215 & SBbc & 0.98 & 0.26 & 40. & 58.9\\ 
 Arp188  &  134.2   &PGC057108 & E-S0 & 0.98 & 0.11 & ... & 49.3\\ 
  & & UGC10214 & Sc & 1.16 & 0.32 & 80.2 & 63.7\\ 
 Arp202  &  47.6   &NGC2719 & I & 1.07 & 0.60 & 131.7 & 90.0\\ 
  & & NGC2719A & I & 0.69 & 0.09 & 150.0 & 39.3\\ 
 Arp205  &  24.7   &UGC06016 & IAB & 1.26 & 0.15 & 45.5 & 50.6\\ 
  & & NGC3448 & S? & 1.47 & 0.53 & 64.8 & 79.2\\ 
 Arp240  &  101.7   &NGC5257 & SABb & 1.17 & 0.28 & 85.1 & 62.1\\ 
  & & NGC5258 & SBb & 1.17 & 0.08 & 177.9 & 34.2\\ 
 Arp242  &  98.2   &NGC4676A & S0-a & 1.34 & 0.23 & 179.1 & 64.4\\ 
  & & NGC4676B & S0-a & 0.99 & 0.17 & 169. & 53.3\\ 
 Arp244  &  24.1   &NGC4038 & SBm & 1.73 & 0.16 & 80 & 51.9\\ 
  & & NGC4039 & SBm & 1.73 & 0.29 & 50 & 71.2\\ 
 Arp245  &  34.0   &NGC2992 & Sa & 1.47 & 0.63 & 17.0 & 90.0\\ 
  & & NGC2993 & Sa & 1.13 & 0.08 & 93.7 & 35.8\\ 
 Arp253  &  28.8   &UGCA173 & SBd & 1.09 & 1.01 & 90.0 & 90.0\\ 
  & & UGCA174 & SBm & 1.12 & 0.53 & 83.1 & 90.0\\ 
 Arp256  &  109.6   &PGC001221 & SBc & 1.04 & 0.46 & 34.8 & 73.3\\ 
  & & PGC001224 & Sb & 0.96 & 0.26 & 98.1 & 60.2\\ 
 Arp261  &  28.7   &PGC052935 & S? & 1.23 & 0.27 & 146.5 & 58.6\\ 
  & & PGC052940 & IB & 1.35 & 0.26 & 148.3 & 66.7\\ 
 Arp269  &  8.5   &NGC4485 & I & 1.30 & 0.18 & 1.7 & 55.2\\ 
  & & NGC4490 & SBcd & 1.83 & 0.61 & 133.1 & 90.0\\ 
 Arp270  &  29.0   &NGC3395 & Sc & 1.20 & 0.26 & 40.5 & 57.8\\ 
  & & NGC3396 & Sm & 1.49 & 0.39 & 97.5 & 90.0\\ 
 Arp271  &  40.0   &NGC5426 & Sc & 1.49 & 0.40 & 0.5 & 69.7\\ 
  & & NGC5427 & SABc & 1.56 & 0.04 & 178. & 25.5\\ 
 Arp279  &  22.6   &NGC1253 & SABc & 1.66 & 0.39 & 84.8 & 68.2\\ 
  & & NGC1253A & SBm & 1.00 & 0.19 & 90.3 & 56.4\\ 
 Arp280  &  14.5   &NGC3769 & Sb & 1.45 & 0.50 & 150.2 & 78.3\\ 
  & & NGC3769A & SBm & 0.96 & 0.39 & 107.0 & 90.0\\ 
 Arp282  &  64.9   &IC1559 & S0-a & 0.92 & 0.28 & 159.4 & 70.3\\ 
  & & NGC0169 & Sab & 1.18 & 0.37 & 92.5 & 69.8\\ 
 Arp283  &  29.6   &NGC2798 & Sa & 1.38 & 0.47 & 160.0 & 84.9\\ 
  & & NGC2799 & SBd & 1.25 & 0.58 & 122.5 & 90.0\\ 
 Arp284  &  38.6   &NGC7714 & Sb & 1.34 & 0.14 & 8.4 & 45.1\\ 
  & & NGC7715 & I & 1.31 & 0.72 & 78.9 & 90.0\\ 
 Arp285  &  44.4   &NGC2854 & SBb & 1.11 & 0.36 & 52.0 & 68.2\\ 
  & & NGC2856 & Sbc & 1.09 & 0.33 & 132.1 & 65.3\\ 
 Arp290  &  46.5   &IC0195 & S0 & 1.16 & 0.30 & 134.8 & 77.3\\ 
  & & IC0196 & Sab & 1.39 & 0.57 & 9.1 & 90.0\\ 
 Arp293  &  82.2   &NGC6285 & S0-a & 1.03 & 0.28 & 110.0 & 68.1\\ 
  & & NGC6286 & Sb & 1.09 & 0.04 & 35. & 90.\\ 
 Arp294  &  43.6   &NGC3786 & SABa & 1.29 & 0.29 & 72.0 & 65.1\\ 
  & & NGC3788 & Sab & 1.30 & 0.54 & 178.8 & 86.0\\ 
 Arp295  &  94.2   &PGC072139 & Sc & 1.27 & 0.58 & 37.5 & 80.0\\ 
  & & PGC072155 & Sb & 1.06 & 0.30 & 103.0 & 63.6\\ 
 Arp297N  &  139.3   &NGC5753 & Sab & 0.73 & 0.10 & 156.0 & 39.3\\ 
  & & NGC5755 & SBcd & 0.66 & 0.17 & 102.5 & 48.1\\ 
 Arp297S  &  70.2   &NGC5752 & Sbc & 0.88 & 0.50 & 121.9 & 79.6\\ 
  & & NGC5754 & SBb & 1.11 & 0.07 & 96.3 & 32.5\\ 
 Arp298  &  66.4   &NGC7469 & Sa & 1.14 & 0.06 & 126.0 & 30.2\\ 
  & & IC5283 & Sc & 1.06 & 0.28 & 105.1 & 60.2\\ 
 NGC2207  &  38.0   &NGC2207 & SABc & 1.69 & 0.25 & 115.9 & 58.2\\ 
  & & IC2163 & Sc & 1.53 & 0.55 & 102.6 & 78.2\\ 
 NGC4567  &  13.9   &NGC4567 & Sbc & 1.44 & 0.10 & 89.0 & 39.4\\ 
  & & NGC4568 & Sbc & 1.63 & 0.36 & 28.6 & 67.5\\ 
\enddata
\tablenotetext{\dag}{All the parameters, except the distance, are extracted from Hyperleda database 
\2014AandA...570A..13M \href{http://leda.univ-lyon1.fr/}{http://leda.univ-lyon1.fr/}.}
\tablenotetext{a}{From the NASA Extragalactic Database (NED), using $H_0 = 73\thinspace\rm{km s^{−1}\thinspace Mpc^{−1}}$, 
with Virgo, Great Attractor, and Shapley Supercluster infall models.}\tablenotetext{b}{Morphological type.
}
\tablenotetext{c}{Log of the length the projected major axis of a galaxy at the isophotal level $25\thinspace \rm{mag/arcsec^2}$ in the B-band, D25 in 0.1 arcmin. 
}
\tablenotetext{d}{Log of the axis ratio of the isophote $25\thinspace \rm{mag/arcsec^2}$ in the  B-band.}
\tablenotetext{e}{Position angle of the major axis of the isophote $25\thinspace \rm{mag/arcsec^2}$ in the B-band (North Eastwards).}
\tablenotetext{f}{Inclination.}
 }}
\end{deluxetable}

\section{NIS galaxies sample}

\begin{deluxetable}{ccccccc}
{\def\arraystretch{0.85}
{

\tabletypesize{\tiny }
\tablecaption{NIS galaxies sample \tablenotemark{\dag}. \label{tab_nissample}}
\tablecolumns{7}
\tablehead{\\
 \colhead{Galaxy} & \colhead{Dist($\rm{Mpc}$)}\tablenotemark{a}  & \colhead{Morph}\tablenotemark{b} & 
 \colhead{logD25}\tablenotemark{c}  &  \colhead{logR25}\tablenotemark{d}  &  
 \colhead{PA($\rm{\deg}$)}\tablenotemark{e}  &   \colhead{i($\rm{\deg}$)}\tablenotemark{f} \\ 
}
\startdata
 NGC24  &  8.2   &Sc & 1.79 & 0.41 & 44.2 & 70.1\\ 
 NGC337  &  22.3   &SBcd & 1.47 & 0.19 & 158 & 50.6\\ 
 NGC628  &  9.9   &Sc & 2.00 & 0.03 & 87. & 19.8\\ 
 NGC925  &  9.3   &Scd & 2.03 & 0.27 & 107.2 & 58.7\\ 
 NGC1097  &  16.5   &SBb & 2.02 & 0.22 & 138.2 & 55.0\\ 
 NGC1291  &  10.1   &S0-a & 2.05 & 0.05 & 156.0 & 29.4\\ 
 NGC2403  &  4.6   &SABc & 2.30 & 0.30 & 126.3 & 61.3\\ 
 NGC2543  &  37.4   &Sb & 1.38 & 0.33 & 52.4 & 66.4\\ 
 NGC2639  &  49.6   &Sa & 1.21 & 0.12 & 140.0 & 44.6\\ 
 NGC2841  &  12.3   &SBb & 1.84 & 0.32 & 147.0 & 65.3\\ 
 NGC2857  &  71.0   &Sc & 1.28 & 0.10 & 90. & 38.0\\ 
 NGC3049  &  24.1   &SBb & 1.32 & 0.24 & 27.8 & 58.0\\ 
 NGC3184  &  10.1   &SABc & 1.87 & 0.01 & 117. & 14.4\\ 
 NGC3344  &  6.9   &Sbc & 1.83 & 0.02 & 150. & 18.7\\ 
 NGC3353  &  18.5   &SABb & 1.13 & 0.14 & 75.7 & 45.5\\ 
 NGC3367  &  47.6   &Sc & 1.46 & 0.01 & 70. & 11.3\\ 
 NGC3521  &  8.0   &SABb & 1.92 & 0.27 & 162.2 & 60.0\\ 
 NGC3621  &  6.5   &SBcd & 1.99 & 0.39 & 161.2 & 67.5\\ 
 NGC3633  &  41.0   &Sa & 1.08 & 0.43 & 70.6 & 78.9\\ 
 NGC3938  &  15.5   &Sc & 1.55 & 0.01 & 28. & 14.1\\ 
 NGC4254  &  39.8   &Sc & 1.70 & 0.03 & 23. & 20.1\\ 
 NGC4321  &  14.1   &SABb & 1.78 & 0.04 & 108. & 23.4\\ 
 NGC4450  &  14.1   &Sab & 1.74 & 0.16 & 173.0 & 48.7\\ 
 NGC4559  &  9.8   &Sc & 2.02 & 0.34 & 148.3 & 64.8\\ 
 NGC4579  &  13.9   &Sb & 1.70 & 0.12 & 90.2 & 41.9\\ 
 NGC4594  &  12.7   &Sa & 1.93 & 0.24 & 89.5 & 59.4\\ 
 NGC4725  &  26.8   &SABa & 1.99 & 0.14 & 35.7 & 45.4\\ 
 NGC4736  &  4.8   &SABa & 1.89 & 0.06 & 105.0 & 31.8\\ 
 NGC4826  &  3.8   &SABa & 2.02 & 0.29 & 114.0 & 64.0\\ 
 NGC5055  &  8.3   &Sbc & 2.07 & 0.22 & 103.0 & 54.9\\ 
 NGC5656  &  51.4   &Sab & 1.10 & 0.15 & 57.2 & 47.6\\ 
 NGC6373  &  51.3   &Sc & 1.01 & 0.17 & 84.2 & 48.9\\ 
 NGC6946  &  5.5   &SABc & 2.06 & 0.02 & 52. & 18.3\\ 
 NGC7331  &  14.4   &Sbc & 1.97 & 0.39 & 169.7 & 70.0\\ 
 NGC7793  &  3.3   &Scd & 2.02 & 0.24 & 89.5 & 63.6\\ 
 UGC04704  &  10.4   &Sd & 1.56 & 0.98 & 115.2 & 90.0\\ 
 UGC05853  &  132.6   &SBc & 1.10 & 0.82 & 36.9 & 90.0\\ 
 UGC06879  &  37.3   &SABc & 1.15 & 0.53 & 167.5 & 75.5\\ 
\enddata
\tablenotetext{\dag}{All the parameters, except the distance, are extracted from Hyperleda database 
\2014AandA...570A..13M \href{http://leda.univ-lyon1.fr/}{http://leda.univ-lyon1.fr/}.}
\tablenotetext{a}{From the NASA Extragalactic Database (NED), using $H_0 = 73\thinspace\rm{km s^{−1}\thinspace Mpc^{−1}}$, 
with Virgo, Great Attractor, and Shapley Supercluster infall models.}\tablenotetext{b}{Morphological type.
}
\tablenotetext{c}{Log of the length the projected major axis of a galaxy at the isophotal level $25\thinspace \rm{mag/arcsec^2}$ in the B-band, D25 in 0.1 arcmin. 
}
\tablenotetext{d}{Log of the axis ratio of the isophote $25\thinspace \rm{mag/arcsec^2}$ in the  B-band.}
\tablenotetext{e}{Position angle of the major axis of the isophote $25\thinspace \rm{mag/arcsec^2}$ in the B-band (North Eastwards).}
\tablenotetext{f}{Inclination.}
 }}
\end{deluxetable}

\section{Clumps photometry table}
\begin{splitdeluxetable*}{ccccccccccBccccccccccccccBcccccccccccccc}


\tabletypesize{\tiny }
\tablecaption{GALEX, {\it Spitzer}, SDSS, H$\alpha$, and 2MASS photometry for the clumps. 
The whole table is available as a machine readable table in the electronic version of the paper and through CDS.\label{tableclumps_photometry}}
\tablecolumns{35}
\tablehead{
\colhead{Name} & \colhead{Ra} & \colhead{Dec} & \colhead{Galaxy} & \colhead{NUV} & \colhead{NUV$_{\rm{err}}$} & \colhead{FUV} & \colhead{FUV$_{\rm{err}}$} & \colhead{$3.6\rm{\mu m}$} & \colhead{$3.6\rm{\mu m}$ $_{\rm{err}}$} & \colhead{$4.5\rm{\mu m}$ } & \colhead{$4.5\rm{\mu m}$ $_{\rm{err}}$} & \colhead{$5.8\rm{\mu m}$ } & \colhead{$5.8\rm{\mu m}$ $_{\rm{err}}$} & \colhead{$8\rm{\mu m}$ } & \colhead{$8\rm{\mu m}$ $_{\rm{err}}$} & \colhead{$24\rm{\mu m}$ } & \colhead{$24\rm{\mu m}$ $_{\rm{err}}$} & \colhead{u} & \colhead{u$_{\rm{err}}$} & \colhead{g} & \colhead{g$_{\rm{err}}$} & \colhead{r} & \colhead{r$_{\rm{err}}$} & \colhead{i} & \colhead{i$_{\rm{err}}$} & \colhead{z} & \colhead{z$_{\rm{err}}$} & \colhead{H$\alpha$+cont} & \colhead{H$\alpha$+cont $_{\rm{err}}$} & \colhead{H$\alpha$} & \colhead{H$\alpha$ $_{\rm{err}}$} & \colhead{H} & \colhead{H$_{\rm{err}}$} & \colhead{J} & \colhead{J$_{\rm{err}}$} & \colhead{K} & \colhead{K$_{\rm{err}}$} \\
\colhead{} & \colhead{$\rm{\deg}$} & \colhead{$\rm{\deg}$} & \colhead{} & \colhead{$\rm{mJy}$}& \colhead{$\rm{mJy}$}& \colhead{$\rm{mJy}$}& \colhead{$\rm{mJy}$}& \colhead{$\rm{mJy}$}& \colhead{$\rm{mJy}$}& \colhead{$\rm{mJy}$}& \colhead{$\rm{mJy}$}& \colhead{$\rm{mJy}$}& \colhead{$\rm{mJy}$}& \colhead{$\rm{mJy}$}& \colhead{$\rm{mJy}$}& \colhead{$\rm{mJy}$}& \colhead{$\rm{mJy}$}& \colhead{$\rm{mJy}$}& \colhead{$\rm{mJy}$}& \colhead{$\rm{mJy}$}& \colhead{$\rm{mJy}$}& \colhead{$\rm{mJy}$}& \colhead{$\rm{mJy}$}& \colhead{$\rm{mJy}$}& \colhead{$\rm{mJy}$}& \colhead{$\rm{mJy}$}& \colhead{$\rm{mJy}$}& \colhead{$\rm{mJy}$}& \colhead{$\rm{mJy}$}& \colhead{$\rm{erg/s/cm^2}$}& \colhead{$\rm{erg/s/cm^2}$}& \colhead{$\rm{mJy}$}& \colhead{$\rm{mJy}$}& \colhead{$\rm{mJy}$}& \colhead{$\rm{mJy}$}& \colhead{$\rm{mJy}$}& \colhead{$\rm{mJy}$} \\
}
\startdata
Arp285\_1\_sbt\_disk\_1\_0 & 141.00604 & 49.204119 & NGC2854 &  0.067 & 0.007  &  0.05 & 0.005  &  0.734 & 0.003  &  0.505 & 0.002  &  2.16 & 0.005  &  5.91 & 0.01  &  12.2 & 0.1  &  0.115 & 0.007  &  0.26 & 0.02  &  0.44 & 0.03  &  0.47 & 0.04  &  0.58 & 0.05  &  0.844 & 0.007  &  2e-14 & 1e-16  &  0.87 & 0.03  &  0.72 & 0.01  &  0.75 & 0.03  \\ 
Arp285\_2\_sbt\_disk\_1\_0 & 141.00945 & 49.206809 & NGC2854 &  0.047 & 0.005  &  0.023 & 0.003  &  0.743 & 0.001  &  0.506 & 0.001  &  1.435 & 0.005  &  3.87 & 0.01  &  11.79 & 0.07  &  0.081 & 0.005  &  0.31 & 0.02  &  0.54 & 0.04  &  0.68 & 0.06  &  0.84 & 0.08  &  0.68 & 0.006  &  1.32e-14 & 1e-16  &  1.49 & 0.05  &  1.06 & 0.02  &  0.68 & 0.03  \\ 
Arp285\_3\_sbt\_disk\_1\_0 & 141.00803 & 49.200117 & NGC2854 &  0.07 & 0.008  &  0.048 & 0.006  &  1.207 & 0.003  &  0.772 & 0.002  &  2.332 & 0.007  &  6.644 & 0.009  &  10.7 & 0.2  &  0.149 & 0.007  &  0.52 & 0.02  &  0.84 & 0.03  &  1.06 & 0.03  &  1.28 & 0.04  &  0.702 & 0.006  &  9.2e-15 & 1e-16  &  1.82 & 0.03  &  1.58 & 0.02  &  1.64 & 0.02  \\ 
Arp285\_4\_sbt\_disk\_1\_0 & 141.01078 & 49.201376 & NGC2854 &  0.099 & 0.006  &  0.08 & 0.005  &  1.75 & 0.02  &  1.12 & 0.01  &  2.88 & 0.04  &  8.6 & 0.1  &  24 & 1  &  0.236 & 0.006  &  0.71 & 0.02  &  1.17 & 0.03  &  1.51 & 0.05  &  1.88 & 0.08  &  0.84 & 0.01  &  8.4e-15 & 2e-16  &  3.71 & 0.05  &  2.86 & 0.04  &  3.14 & 0.04  \\ 
Arp285\_5\_sbt\_disk\_1\_0 & 141.02216 & 49.204905 & NGC2854 &  0.044 & 0.003  &  0.042 & 0.002  &  0.325 & 0.0008  &  0.209 & 0.002  &  0.476 & 0.006  &  1.441 & 0.006  &  2.21 & 0.09  &  0.064 & 0.002  &  0.213 & 0.006  &  0.3 & 0.01  &  0.33 & 0.01  &  0.3 & 0.02  &  0.218 & 0.006  &  3e-15 & 1e-16  &  0.25 & 0.03  &  0.3 & 0.02  &  0.26 & 0.03  \\ 
Arp285\_6\_sbt\_disk\_1\_0 & 141.0653 & 49.250892 & NGC2856 &  0.158 & 0.006  &  0.102 & 0.005  &  8.18 & 0.01  &  5.637 & 0.007  &  18.97 & 0.02  &  56.22 & 0.05  &  186.2 & 0.7  &  0.521 & 0.008  &  1.65 & 0.02  &  3.26 & 0.05  &  4.47 & 0.07  &  6.1 & 0.1  &  3.44 & 0.01  &  5.81e-14 & 1e-16  &  12.36 & 0.07  &  9.32 & 0.04  &  10.22 & 0.08  \\ 
Arp285\_7\_sbt\_disk\_1\_0 & 141.06919 & 49.247372 & NGC2856 &  0.093 & 0.003  &  0.048 & 0.002  &  8.79 & 0.03  &  6.12 & 0.02  &  22.21 & 0.02  &  65.37 & 0.04  &  191.3 & 0.9  &  0.451 & 0.007  &  1.31 & 0.02  &  2.84 & 0.05  &  3.89 & 0.07  &  5.6 & 0.1  &  4.46 & 0.01  &  9.66e-14 & 1e-16  &  11.7 & 0.1  &  8.8 & 0.08  &  10.45 & 0.08  \\ 
Arp285\_8\_sbt\_disk\_1\_0 & 141.07527 & 49.241805 & NGC2856 &  0.0032  & <  &  0.0019  & <  &  0.126 & 0.001  &  0.083 & 0.001  &  0.118 & 0.004  &  0.299 & 0.005  &  0.74 & 0.05  &  0.012  & <  &  0.045 & 0.003  &  0.091 & 0.006  &  0.118 & 0.008  &  0.13 & 0.01  &  0.048  & <  &  2.1e-16  & <  &  0.17  & <  &  0.3 & 0.02  &  0.24  & <  \\ 
\enddata
\end{splitdeluxetable*}

\section{CIGALE output parameters of the clumps}
\begin{deluxetable*}{cccccccccccc}


\tabletypesize{\tiny }
\tablecaption{Output parameters of CIGALE. 
The whole table is available as a machine readable table in the electronic version of the paper and through CDS.\label{tab_cig_out}}
\tablecolumns{12}
\tablehead{
\colhead{Name} &  \colhead{SFR} & \colhead{SFR$_{\rm{err}}$} & \colhead{$\log M_{*}$} & \colhead{$\log M_{*\thinspace \rm{err}}$} & \colhead{$\rm{age}_{\rm{burst}}$} & \colhead{$\rm{age}_{\rm{burst}\thinspace \rm{err}}$} & \colhead{$f_{\rm{burst}}$ } & \colhead{$f_{\rm{burst}\thinspace \rm{err}}$} & \colhead{$A_{\rm{H\alpha}}$ }  & \colhead{$A_{\rm{FUV}}$ } & \colhead{$\chi^{2}_{\rm{red}}$}   \\
\colhead{} & \colhead{$\rm{M_{\sun}}/\rm{yr}$} & \colhead{$\rm{M_{\sun}}/\rm{yr}$} & \colhead{$\rm{M_{\sun}}$}& \colhead{$\rm{M_{\sun}}$}& \colhead{$\rm{Myr}$}& \colhead{$\rm{Myr}$}& \colhead{}& \colhead{}&  \colhead{mag}& \colhead{mag}& \colhead{} \\
}
\startdata
Arp285\_1\_sbt\_disk\_1\_0 &  0.06 & 0.02  &  7.6 & 0.5  &  120 & 40  &  0.6 & 0.3  &  2.1  &  3.0  &  1.4  \\ 
Arp285\_2\_sbt\_disk\_1\_0 &  0.029 & 0.007  &  8.1 & 0.3  &  310 & 60  &  0.4 & 0.3  &  2.3  &  3.2  &  2.6  \\ 
Arp285\_3\_sbt\_disk\_1\_0 &  0.003 & 0.003  &  8.3 & 0.3  &  170 & 30  &  0.3 & 0.2  &  2.4  &  2.8  &  1.2  \\ 
Arp285\_4\_sbt\_disk\_1\_0 &  0.005 & 0.004  &  8.5 & 0.2  &  90 & 10  &  0.23 & 0.09  &  2.4  &  2.8  &  1.8  \\ 
Arp285\_5\_sbt\_disk\_1\_0 &  0.001 & 0.001  &  7.4 & 0.4  &  130 & 30  &  0.6 & 0.3  &  1.5  &  1.9  &  4.4  \\ 
Arp285\_6\_sbt\_disk\_1\_0 &  0.61 & 0.06  &  9.0 & 0.2  &  220 & 30  &  0.3 & 0.2  &  3.1  &  4.1  &  2.1  \\ 
Arp285\_7\_sbt\_disk\_1\_0 &  0.32 & 0.02  &  9.17 & 0.05  &  300 & 10  &  0.2 & 0.01  &  2.9  &  4.0  &  5.6  \\ 
Arp285\_8\_sbt\_disk\_1\_0 &  0.0022 & 0.0009  &  7.8 & 0.2  &  200 & 200  &  0.1 & 0.2  &  1.4  &  3.1  &  0.61  \\ 
\enddata
\end{deluxetable*}

\section{Integrated photometry table}
\begin{splitdeluxetable*}{ccccccccccBccccccccccccccBcccccccccccc}


\tabletypesize{\tiny }
\tablecaption{GALEX, {\it Spitzer}, SDSS, H$\alpha$, and 2MASS photometry for the SB\&T and NIS galaxies. Total fluxes have not been corrected for Galactic absorption. 
The whole table is available as a machine readable table in the electronic version of the paper and through CDS.\label{tableintegrated_photometry}}
\tablecolumns{33}
\tablehead{
\colhead{System} & \colhead{Galaxy} & \colhead{NUV} & \colhead{NUV$_{\rm{err}}$} & \colhead{FUV} & \colhead{FUV$_{\rm{err}}$} & \colhead{$3.6\rm{\mu m}$} & \colhead{$3.6\rm{\mu m}$ $_{\rm{err}}$} & \colhead{$4.5\rm{\mu m}$ } & \colhead{$4.5\rm{\mu m}$ $_{\rm{err}}$} & \colhead{$5.8\rm{\mu m}$ } & \colhead{$5.8\rm{\mu m}$ $_{\rm{err}}$} & \colhead{$8\rm{\mu m}$ } & \colhead{$8\rm{\mu m}$ $_{\rm{err}}$} & \colhead{$24\rm{\mu m}$ } & \colhead{$24\rm{\mu m}$ $_{\rm{err}}$} & \colhead{u} & \colhead{u$_{\rm{err}}$} & \colhead{g} & \colhead{g$_{\rm{err}}$} & \colhead{r} & \colhead{r$_{\rm{err}}$} & \colhead{i} & \colhead{i$_{\rm{err}}$} & \colhead{z} & \colhead{z$_{\rm{err}}$} & \colhead{H$\alpha$+cont} & \colhead{H$\alpha$+cont $_{\rm{err}}$} & \colhead{H$\alpha$} & \colhead{H$\alpha$ $_{\rm{err}}$} & \colhead{H} & \colhead{H$_{\rm{err}}$} & \colhead{J} & \colhead{J$_{\rm{err}}$} & \colhead{K} & \colhead{K$_{\rm{err}}$} \\
\colhead{}  & \colhead{} & \colhead{$\rm{mJy}$}& \colhead{$\rm{mJy}$}& \colhead{$\rm{mJy}$}& \colhead{$\rm{mJy}$}& \colhead{$\rm{mJy}$}& \colhead{$\rm{mJy}$}& \colhead{$\rm{mJy}$}& \colhead{$\rm{mJy}$}& \colhead{$\rm{mJy}$}& \colhead{$\rm{mJy}$}& \colhead{$\rm{mJy}$}& \colhead{$\rm{mJy}$}& \colhead{$\rm{mJy}$}& \colhead{$\rm{mJy}$}& \colhead{$\rm{mJy}$}& \colhead{$\rm{mJy}$}& \colhead{$\rm{mJy}$}& \colhead{$\rm{mJy}$}& \colhead{$\rm{mJy}$}& \colhead{$\rm{mJy}$}& \colhead{$\rm{mJy}$}& \colhead{$\rm{mJy}$}& \colhead{$\rm{mJy}$}& \colhead{$\rm{mJy}$}& \colhead{$\rm{mJy}$}& \colhead{$\rm{mJy}$}& \colhead{$\rm{erg/s/cm^2}$}& \colhead{$\rm{erg/s/cm^2}$}& \colhead{$\rm{mJy}$}& \colhead{$\rm{mJy}$}& \colhead{$\rm{mJy}$}& \colhead{$\rm{mJy}$}& \colhead{$\rm{mJy}$}& \colhead{$\rm{mJy}$} \\
}
\startdata
Arp120 & NGC4438 &  4.346 & 0.002  &  2.337 & 0.003  &  510.71 & 0.03  &  301.13 & 0.04  &  339.8 & 0.1  &  430 & 20  & ... & ...  &  36.28 & 0.04  &  166.29 & 0.02  &  334.43 & 0.05  &  506.5 & 0.03  &  648.7 & 0.2  &  352.6 & 0.4  &  5.68e-13 & 2e-15  &  1205.7 & 0.4  &  969.7 & 0.3  &  970.5 & 0.4  \\ 
Arp120 & NGC4435 &  0.8707 & 0.0006  &  0.1615 & 0.0009  &  234.41 & 0.01  &  139.42 & 0.01  &  111.54 & 0.04  &  143 & 3  &  109.016 & 0.005  &  16.5 & 0.01  &  84.759 & 0.006  &  167.66 & 0.02  &  254.206 & 0.009  &  325.41 & 0.05  &  215.6 & 0.2  &  7.6e-14 & 1e-15  &  590.3 & 0.2  &  478.45 & 0.09  &  470.6 & 0.2  \\ 
Arp178 & NGC5614 &  1.525 & 0.003  & ... & ...  &  173.71 & 0.02  &  98.59 & 0.02  &  107.57 & 0.09  &  225 & 8  &  191.55 & 0.03  &  12.33 & 0.04  &  53.88 & 0.01  &  111.3 & 0.04  &  165.3 & 0.02  &  210.8 & 0.2  &  129.2 & 0.4  &  1.08e-13 & 3e-15  &  429.6 & 0.5  &  309.7 & 0.3  &  299.4 & 0.6  \\ 
Arp178 & NGC5613 &  0.1235 & 0.0006  & ... & ...  &  7.909 & 0.004  &  4.632 & 0.004  &  5.4 & 0.02  &  6.62 & 0.03  &  4.048 & 0.007  &  0.691 & 0.009  &  2.772 & 0.003  &  5.45 & 0.01  &  7.779 & 0.005  &  10.01 & 0.04  &  5.7 & 0.1  &  2e-15 & 7e-16  &  20.0 & 0.1  &  15.14 & 0.08  &  12.3 & 0.2  \\ 
Arp181 & NGC3215 &  0.581 & 0.001  &  0.297 & 0.001  &  27.121 & 0.004  &  17.498 & 0.004  &  21.68 & 0.02  &  50.3 & 0.1  &  34.098 & 0.004  &  2.8 & 0.01  &  10.928 & 0.004  &  20.93 & 0.01  &  29.013 & 0.007  &  36.33 & 0.05  &  14.8 & 0.09  &  2.7e-14 & 5e-16  &  62.6 & 0.2  &  50.4 & 0.1  &  53.5 & 0.2  \\ 
Arp181 & NGC3212 &  0.4046 & 0.0009  &  0.1702 & 0.0008  &  15.096 & 0.003  &  10.416 & 0.003  &  22.81 & 0.01  &  71.9 & 0.4  &  96.703 & 0.003  &  1.59 & 0.01  &  5.953 & 0.003  &  10.02 & 0.01  &  13.428 & 0.006  &  16.56 & 0.04  &  6.71 & 0.07  &  8.6e-15 & 4e-16  &  35.4 & 0.2  &  24.27 & 0.09  &  27.6 & 0.1  \\ 
\enddata
\end{splitdeluxetable*}



\end{document}